\documentclass[pdflatex,sn-mathphys-num]{sn-jnl}

\usepackage{graphicx}%
\usepackage{multirow}%
\usepackage{amsmath,amssymb,amsfonts}%
\usepackage{amsthm}%
\usepackage{mathrsfs}%
\usepackage[title]{appendix}%
\usepackage{xcolor}%
\usepackage{textcomp}%
\usepackage{manyfoot}%
\usepackage{booktabs}%
\usepackage{algorithm}%
\usepackage{algorithmicx}%
\usepackage{algpseudocode}%
\usepackage{listings}%
\UseRawInputEncoding
\usepackage{float}
\usepackage{booktabs}
\usepackage{tabularx}
\usepackage{makecell}
\usepackage{pdflscape}

\theoremstyle{thmstyleone}%
\theoremstyle{thmstyletwo}%

\theoremstyle{thmstylethree}%

\begin{document}

\title[Article Title]{Data-Driven Discovery of a Hyperelastic Strain-Energy Function for Veins Using Constitutive Artificial Neural Networks}

\author[1]{\fnm{Marisa} \sur{Bazzi}}\email{msbazzi@stanford.edu}
\equalcont{These authors contributed equally to this work.}
\author[2]{\fnm{Divya} \sur{Adil}}\email{divyaad@stanford.edu}
\equalcont{These authors contributed equally to this work.}
\author[3]{\fnm{Gavin} \sur{Mays}}\email{gavin.mays@yale.edu}
\author[2,4]{\fnm{Ellen} \sur{Kuhl}} \email{ekuhl@stanford.edu}
\author[5,6]{\fnm{Christopher K.} \sur{Breuer}} \email{Christopher.Breuer@nationwidechildrens.org}
\author[3]{\fnm{Jay D.} \sur{Humphrey}}\email{jay.humphrey@yale.edu}
\author*[1,2,4]{\fnm{Alison L.} \sur{Marsden}}\email{amarsden@stanford.edu}

\affil[1]{\orgdiv{Department of Pediatrics}, \orgname{Stanford University}, \orgaddress{\city{Stanford},  \state{CA}, \country{United States}}}

\affil[2]{\orgdiv{Department of Mechanical Engineering}, \orgname{Stanford University}, \orgaddress{\city{Stanford}, \state{CA}, \country{United States}}}

\affil[3]{\orgdiv{Department of Biomedical Engineering}, \orgname{Yale University}, \orgaddress{\city{New Haven}, \state{CT}, \country{United States}}}

\affil[4]{\orgdiv{Department of Bioengineering}, \orgname{Stanford University}, \orgaddress{ \city{Stanford}, \state{CA}, \country{United States}}}

\affil[5]{\orgdiv{Center for Regenerative Medicine}, \orgname{Nationwide Children’s Hospital}, \orgaddress{\city{Columbus}, \state{OH}, \country{United States}}}

\affil[6]{\orgdiv{Department of Biomedical Engineering}, \orgname{The Ohio State University}, \orgaddress{ \city{Columbus}, \state{OH}, \country{United States}}}


\abstract{
Veins are commonly subjected to altered mechanical loading in clinical settings such as vein grafting, arteriovenous fistula creation, and tissue-engineered vascular graft implantation, making accurate constitutive models of venous tissue clinically important. Venous mechanical behavior differs characteristically from that of arteries, and existing constitutive models struggle to simultaneously fit the circumferential and axial stress-stretch responses across vein types.
Here, we present a constitutive artificial neural network (CANN) framework that discovers a compact, interpretable hyperelastic constitutive model from biaxial mechanical data across three types of veins (saphenous, jugular, and inferior vena cava). The framework incorporates an expanded candidate function library with structurally motivated terms representing progressive fiber recruitment and fiber-family interactions. Sobol sensitivity analysis, correlation-based pruning, and response-level family collapse were then used to reduce the discovered basis to a compact strain-energy density function (SEF). The resulting eight-term SEF achieved significantly higher fitting accuracy than established venous constitutive models across all three types of veins $(R^2>0.9, p<0.01)$.
These findings establish a framework for constitutive model discovery in veins and propose a candidate constitutive form for further validation across larger, independent cohorts.}




\maketitle

\section{Introduction}\label{intro}
Veins are the primary capacitance vessels of the circulation, storing approximately 60--80\% of total blood volume depending on posture and activity~\cite{desai_inhibition_2010}. Clinical interventions, including coronary artery bypass grafting (CABG), peripheral bypass surgery, Fontan completion, arteriovenous fistulas, and tissue-engineered vascular graft (TEVG) implantation, can expose veins to dramatically altered biomechanical loading. The clinical consequences of this altered mechanical environment are particularly evident when an autologous saphenous vein is used as an arterial bypass grafts, with reported failure rates of 30--50\% at 5 years following arterial reconstruction~\cite{owens_lower_2008}. More broadly, the long-term performance of small-diameter grafts remains limited by intimal hyperplasia, stenosis, and subsequent atherosclerotic changes \cite{greenwald_improving_2000}. A broad body of experimental and clinical work indicates that such interventions abruptly expose veins to non-native hemodynamic and mechanical conditions, including altered pressure, circumferential stretch, axial loading, and flow. These changes can initiate growth and remodeling responses that may become maladaptive, and ultimately contribute to failure. In TEVG, differences in material properties between the host vessel and implanted graft may further amplify local perturbations in wall stress and flow. Reduced venous compliance has likewise been associated with early failure of an arteriovenous fistula. ~\cite{prim_evaluation_2021,salacinski_mechanical_2001, abbott_effect_1987, yao_vivo_2023, ballyk_compliance_1997,stewart_effects_1992, smith_impact_2016}. Together, these observations underscore the need for constitutive models that can accurately represent vein mechanics under physiological, pathophysiological, and surgically altered loading conditions.

Despite this need, constitutive modeling of the venous wall remains comparatively underdeveloped relative to arterial modeling. This disparity is important because venous and arterial tissues differ substantially in composition, organization, and mechanical function \cite{mattson_structural_2017,vessels_comparison_Ramachandra2024-je}. Veins generally have a higher collagen-to-elastin ratio than arteries. In the vena cava, for example, the collagen content is nearly four times the elastin content, whereas arteries exhibit a more balanced composition that varies with location along the vascular tree. Venous elastin tends to be predominantly oriented in the axial direction, while the collagen is more circumferentially organized and straightens at relatively higher strains ~\cite{mattson_structural_2017}. Compared to arteries, these structural differences produce a distinct mechanical response characterized by a relatively flat toe-region and sudden stiffening behavior in the circumferential direction. Consequently, constitutive models developed primarily for arteries need do not capture the characteristic mechanical behavior of veins.

Nevertheless, many constitutive models that have been applied to veins in prior work were adapted from arterial formulations, ranging from phenomenological descriptors of the overall tissue response to structurally motivated models that represent the contributions of the ground matrix and fiber families \cite{kao_constitutive_2010, anssari-benam_specialized_2020, saez_microstructural_2016, sokolis_experimental_2013, humphrey_cardiovascular_2002}. For example, a four-fiber family model was able to distinguish mechanical differences across different types of vein but did not accurately capture the measured stress-stretch response from biaxial mechanical data \cite{Lee2013-ao}. Similarly, the phenomenological model of Desch et al. \cite{Desch2007-is} and the structural model of Rezakhaniha et al. \cite{Rezakhaniha2008-nh} exhibited limited fitting accuracy. Improved agreement for porcine jugular veins was obtained only after introducing a quadratic circumferential--axial coupling term, suggesting that formulations without explicit cross-directional coupling may not adequately capture interactions between the principal loading directions~\cite{sokolis_experimental_2013}. Other studies have shown that richer anisotropic descriptions for elastin improved accuracy for three-dimensional characterization under distension-extension loading, whereas isotropic elastin formulations led to substantially poorer fits to pressure-radius and pressure-axial force data ~\cite{Rezakhaniha2008-nh, saitta-rezakhaniha_biomechanics_2010}. Despite these advances, existing formulations still struggle to fit the axial and circumferential responses simultaneously, and prior models have not generalized well across the different typesof veins. This persistent gap motivates development of a constitutive model for veins that incorporates the necessary directional interactions while remaining generalizable across subjects and vein types.

Constitutive artificial neural networks (CANNs) provide a systematic framework for identifying mechanistically informed constitutive models directly from experimental data. By embedding continuum-mechanics principles like material frame indifference and thermodynamic consistency into the network architecture, CANNs learn an interpretable strain-energy density function (SEF) rather than directly approximating stress through a black-box neural network ~\cite{linka_constitutive_2021,Linka2023-ka_CANN_constraints,CANN_Linka2022-ak,pierre_principal-stretch-based_2023, peirlinck_universal_2024}. CANNs have been successfully applied to discover improved constitutive models in a wide range of applications, including human brain tissue \cite{pierre_principal-stretch-based_2023} and arterial aging and disease progression \cite{CANN_Linka2022-ak}. However, its discovery capability is restricted by the candidate library: features not present cannot be discovered. Existing CANN libraries considered first in this study primarily comprised invariant-based terms adapted from arterial constitutive models, offering limited representation of directional coupling, fiber recruitment, or richer higher-order strain functions.

The distinct microstructure and nonlinear response of veins motivate expanding the candidate library for CANN-based constitutive discovery beyond what is commonly used for arteries. Progressive recruitment of initially crimped collagen fibers provides a structural explanation for the flat toe region followed by abrupt stiffening in the venous circumferential response~\cite{Famaey2015-vk_review_of_bridging_veins}. Similar transitions from compliant low-strain behavior to rapid stiffening have been modeled in other collagen-rich tissues using distributions of fiber slack stretches ~\cite{li_multi-scale_2023}. Interactions among directional tissue constituents may also contribute to the coupled axial and circumferential behavior of veins. These considerations motivated a model-discovery framework capable of evaluating multiple constitutive terms without specifying which would dominate the venous response.

Here, we present a CANN framework that discovers a compact, interpretable constitutive model for veins from an expanded candidate library. The CANN was trained on three types of ovine vein samples to identify candidate terms supported by all vein types, yielding a discovered SEF. The terms in the discovered SEF were systematically ranked, collapsed, and refitted to derive a reduced closed-form SEF. The mechanical roles of the retained terms were evaluated through ablation analysis, and the resulting reduced model was compared with established constitutive formulations based on its overall fitting accuracy and model complexity.

\section{Methods}\label{methods}
\begin{figure}[!htbp]
\centering
\includegraphics[width=1.0\textwidth]{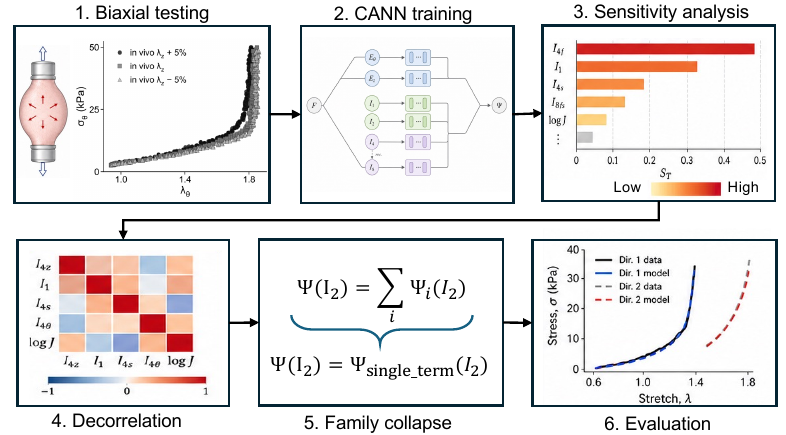}
\caption{
Workflow for CANN-based constitutive model discovery in venous tissue. 
(1) Biaxial pressure--diameter and force--length data were converted to stress--stretch responses and used to (2) train animal-specific CANNs with strain-based, invariant-based, and recruitment-modified branches. 
Surviving retained terms were (3) ranked using Sobol sensitivity analysis, (4) filtered using correlation-based redundancy reduction, and compacted using a (5) response-based within-family collapse to obtain a reduced closed-form strain-energy density function. 
The resulting model was (6) evaluated against the experimental biaxial data and compared with established constitutive models from the literature.}
\label{fig:schema}
\end{figure}

Our overall workflow for constitutive model discovery for veins consists of six main stages: (1) experimental biaxial characterization, (2) subject-specific CANN training, (3) post-training sensitivity ranking, (4) correlation-based pruning, (5) within-family reduction, and (6) reduced-model evaluation. Pressure--diameter and axial force--length data were converted into circumferential and axial stretches and stresses under the standard assumptions of axisymmetric, incompressible deformation. Subject-specific CANNs were then trained using a broad library of invariant-based, strain-based, coupling, and recruitment-modified candidate energy terms, and the retained terms in the resulting discovered SEF were ranked using Sobol sensitivity analysis. Correlation-based redundancy filtering and response-based within-family collapse were subsequently applied to identify a compact and interpretable reduced candidate basis, which was systematically refined and refitted to obtain a final reduced SEF describing biaxial venous mechanics across all cohorts. The reduced model was then evaluated through nested and targeted ablation analyses to determine the contribution of each retained constitutive mechanism, and compared with established constitutive formulations in terms of global fitting accuracy and information criteria. A schematic of the workflow is shown in Fig.~\ref{fig:schema}; the following sections describe each component in detail.

\subsection{Vein specimens and biaxial testing}
The study included nine ovine venous specimens spanning three types: three saphenous, three inferior vena cava, and three jugular (Table~\ref{tab:training_points_by_vein}), chosen to evaluate the model discovery framework across a range of mechanical responses rather than within a single vessel type. The saphenous vein samples were obtained from polypay cross sheep (n=3, 39-46 kg, 5 months old). The IVC samples were from two female and one male healthy sheep (n=3, 36-37 kg, 6-7 months old). The jugular vein samples were obtained from healthy adult male Dorset sheep (n=3, 45-55 kg, 7-8 months old)\cite{jugular_vein_dataset_Ramachandra2022-jy}. Tubular samples from each specimen were mechanically tested using cyclic distension--extension biaxial testing~\cite{Ferruzzi2013-mg_exp_method}. Measured quantities included luminal pressure, axial force, axial length, and outer radius. Pressure--diameter (P-d) tests were performed at fixed axial stretches corresponding to the estimated in vivo value and $\pm 5\%$ of this value. Force--length (f-l) tests were performed at fixed luminal pressures ranging from $2-30$ mmHg.

Circumferential and axial stretches and Cauchy stresses were computed from the measured pressure, axial force, axial length, and inner and outer radii under standard assumptions of radially-averaged, axisymmetric, incompressible deformation, following the approach of Ferruzzi et al.~\cite{Ferruzzi2013-mg_exp_method}. The resulting paired quantities $(\lambda_\theta,\lambda_z,\sigma_\theta,\sigma_z)$ were used as inputs for subject-specific CANN training.

\begin{figure}[!htbp]
\centering
\includegraphics[width=0.85\textwidth]{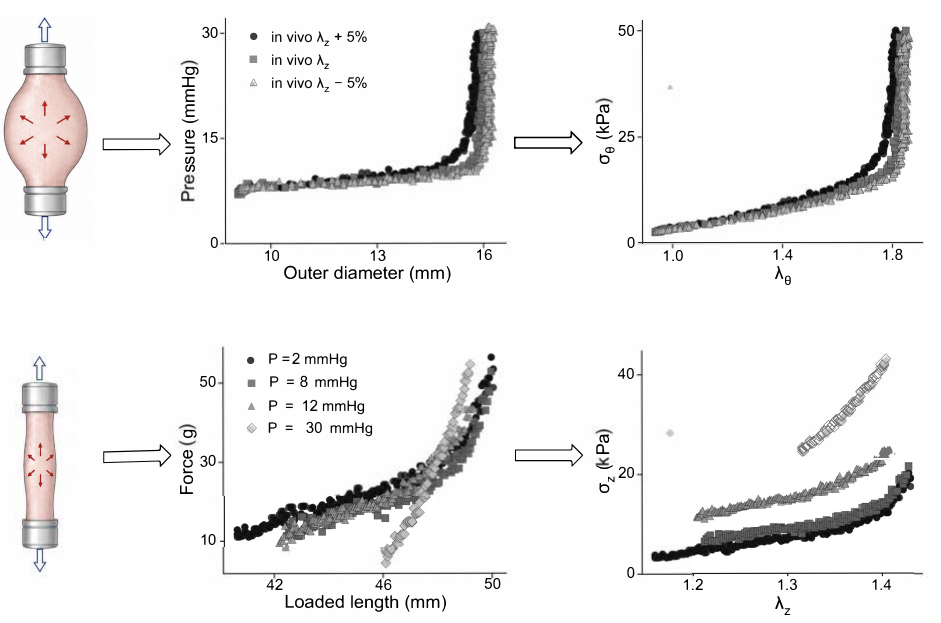}
\caption {Illustrative distension-extension biaxial testing. 
Cylindrical vein samples were subjected to pressure-diameter protocols at multiple values of fixed axial stretch and, force-length protocols at fixed luminal pressure. 
Measured pressure, axial force, axial length, and outer diameter were converted to circumferential and axial stretches and radially averaged Cauchy stresses under assumptions of homogeneous, axisymmetric, incompressible deformation.}
\label{fig:biaxial_mechanical_testing}
\end{figure}

\begin{table}[htbp]
\centering
\small
\caption{Stress--stretch data sets used for CANN training, separated by vein type. Each sample was tested under three pressure--diameter (P-d) protocols at $\lambda_z^{iv}{-}5\%$, $\lambda_z^{iv}$, and $\lambda_z^{iv}{+}5\%$, and four force--length (f-l) protocols at 2 mmHg, 8 mmHg, 12 mmHg, and 30 mmHg. Point counts represent all admissible $(\lambda,\sigma)$ pairs.}
\label{tab:training_points_by_vein}
\begin{tabular*}{\textwidth}{@{\extracolsep{\fill}}lcccccc}
\toprule
\multirow{2}{*}{Vein type} & \multirow{2}{*}{Samples} & \multicolumn{2}{c}{P-d} & \multicolumn{2}{c}{f-l} & \multirow{2}{*}{Total points} \\
\cmidrule(lr){3-4}\cmidrule(lr){5-6}
 &  & Curves & Points & Curves & Points &  \\
\midrule
Saphenous vein          & 3 & 9 & 6\,168 & 12 & 3\,575 & 9\,743 \\
Inferior vena cava      & 3 &  9 & 3\,901 & 12 & 3\,156 & 7\,057 \\
Jugular vein \cite{jugular_vein_dataset_Ramachandra2022-jy} & 3 & 9 & 3\,930 & 12 & 3\,234 & 7\,164 \\
\midrule
\multicolumn{2}{l}{Total} & 27 & 13\,999 & 36 & 9\,965 & 23\,964 \\
\botrule
\end{tabular*}
\end{table}

\subsection{Constitutive artificial neural network}
\subsubsection{Architecture}
The CANN architecture builds on established constitutive-discovery frameworks \cite{linka_constitutive_2021, peirlinck_universal_2024}, encoding prior knowledge from continuum mechanics into the network at three levels: material frame indifference and incompressibility are enforced through the choice of input deformation tensors, while additional physical constraints on the SEF are imposed through activation functions satisfying non-negativity, appropriate growth conditions, and being infinite at extreme deformation. Thermodynamic consistency is ensured by formulating the network output as the SEF and obtaining the stresses through differentiation. 

\begin{figure}[!htbp]
\centering
\includegraphics[width=1.0\textwidth]{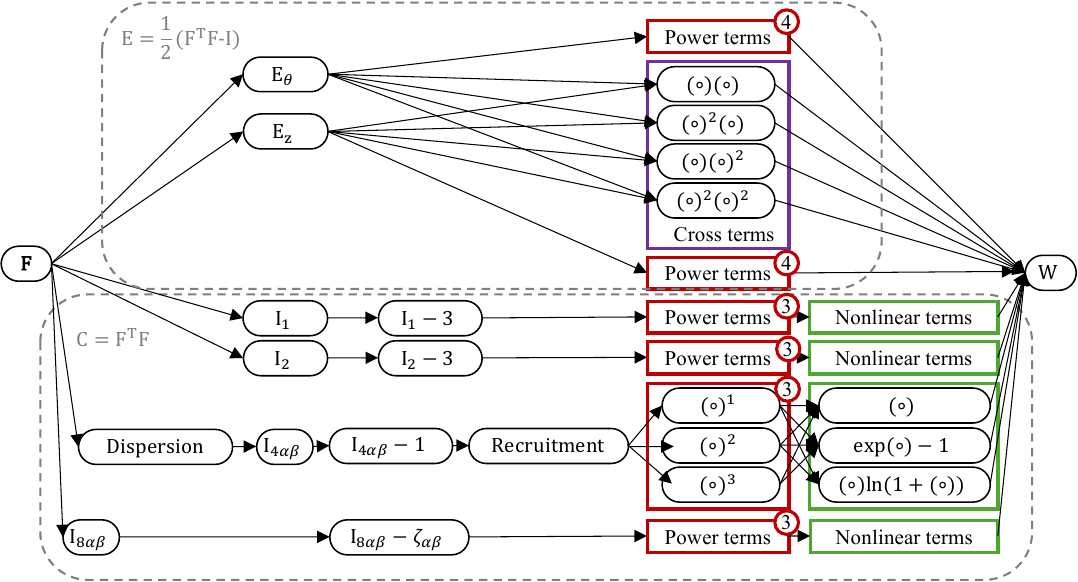}
\caption{
Architecture of the constitutive artificial neural network constructed for veins. The figure shows the input deformation gradient, $\mathbf{F}$, splitting into branches based on the Green-Lagrange strain $\mathbf{E}$ and invariants of the right Cauchy-Green tensor $\mathbf{C}$, before going through layers of activation functions. The weighted outputs of all branches are summed to obtain the output strain-energy density function, $W$.
}\label{fig:network}
\end{figure}

Figure \ref{fig:network} shows the architecture of our CANN for veins. The input deformation gradient $\mathbf{F}$ and right Cauchy--Green tensor $\mathbf{C}$ were calculated from the experimental data as
\begin{equation}
    \mathbf{F}
    = \lambda_\theta \mathbf{e}_\theta \otimes \mathbf{E}_\theta
    + \lambda_z \mathbf{e}_z \otimes \mathbf{E}_z
    + \lambda_r \mathbf{e}_r \otimes \mathbf{E}_r ,
    \qquad
    \mathbf{C}
    = \lambda_\theta^2 \mathbf{E}_\theta \otimes \mathbf{E}_\theta
    + \lambda_z^2 \mathbf{E}_z \otimes \mathbf{E}_z
    + \lambda_r^2 \mathbf{E}_r \otimes \mathbf{E}_r ,
\end{equation}
where $\lambda_r$, $\lambda_\theta$, and $\lambda_z$ are the mean radial, circumferential, and axial stretches, respectively, and the incompressibility constraint is enforced through
\begin{equation}
    \lambda_r = \frac{1}{\lambda_\theta \lambda_z}.
\end{equation}
The neural network consists of two main sets of candidate branches: invariant-based and strain-based. The invariant branch consists of isotropic and anisotropic terms. The first two invariants of $\mathbf{C}$,
\begin{equation}
    I_1 = \mathrm{tr}(\mathbf{C}),
    \qquad
    I_2 =\frac{1}{2}\left[ I_1^2 - \mathrm{tr}(\mathbf{C}^2)
    \right]
\end{equation}
constitute the isotropic part. Anisotropic branches represent fiber behaviors through the fourth and eighth invariants. The fourth invariants with dispersion \cite{hgo_model_Gasser2006-vd} are defined as 
\begin{equation}
    I_{4i} = \mathbf{C}:\mathbf{H}_i,
\end{equation}
where $\mathbf{H}_i$ is a structural tensor associated with fiber family $i$, defined as
\begin{equation}
    \mathbf{H}_i  =  \kappa_i \mathbf{I}
    + (1-3\kappa_i)\mathbf{d}_i\otimes\mathbf{d}_i,
    \qquad
    0 \leq \kappa_i \leq \frac{1}{3}.
\end{equation}
here $\kappa_i=0$ represents perfectly aligned fibers, whereas $\kappa_i=1/3$ represents an isotropic orientation distribution. Four fiber families were considered: circumferential, axial, and two diagonal families. The diagonal families were defined by
\begin{equation}
    \mathbf{d}_{1,2} =  \cos\phi_{1,2}\,\mathbf{E}_\theta   +  \sin\phi_{1,2}\,\mathbf{E}_z ,
\end{equation}
where $\phi_{1,2}$ are the fiber orientation angles of the two diagonal fiber families measured from the circumferential direction. To better capture the long toe region characteristic of venous tissue, we modified the $I_{4i}$ invariants to incorporate fiber recruitment (\ref{fiber-recruit}).

The eighth invariant representing coupling between the two diagonal fiber families was also included,
\begin{equation}
    I_{8d_1d_2} =  \mathbf{d}_1 \cdot \mathbf{C}\mathbf{d}_2 .
\end{equation}
In parallel, the strain-based branches are functions of the Green--Lagrange strain components in the circumferential and axial directions,
\begin{equation}
    E_\theta = \frac{1}{2}(\lambda_\theta^2-1), \qquad  E_z =  \frac{1}{2}(\lambda_z^2-1).
\end{equation}
Each shifted invariant terms pass through a polynomial power layer and a non-linear layer with exponential and logarithmic functions. The strain terms also pass through a polynomial layer, with the addition of higher-order cross terms (see ~\ref{fig:network}).

The invariant and strain candidate outputs are concatenated into a single shared feature vector and passed through a final non-negative linear layer to get the SEF with the general form
\begin{equation}
    W(\mathbf{C}, \mathbf{\eta}) = \sum_{j=1}^{N_b}  c_j \psi_j(\mathbf{C}, \mathbf{\eta}_j),  \qquad c_j \geq 0,
\end{equation}
where $\psi_j$ denotes one candidate basis function from the invariant or strain library, $c_j$ is the learned non-negative coefficient, $N_b$ is the total number of candidate terms, and $\mathbf{\eta}_j$ contains any nonlinear parameters associated with candidate $j$ . This additive structure follows the CANN concept that each constitutive neuron represents one candidate contribution to the strain-energy density function, with the network assigning a nonnegative weight to each candidate contributions during training ~\cite{linka_constitutive_2021, peirlinck_universal_2024}. More details on loss function and training can be found in \ref{loss}.

\subsubsection{Fiber recruitment}\label{fiber-recruit}
Motivated by previous work on collagen-rich tissue \cite{Famaey2015-vk_review_of_bridging_veins,li_multi-scale_2023}, we expanded the candidate library to include recruitment-modified fiber terms alongside conventional fiber contributions. This allowed the model discovery to determine whether recruitment of initially crimped fibers was supported by the biaxial data.

For each fiber family $i$, the effective fiber-family stretch was defined from the $I_4$ invariant as
\begin{equation}
    \lambda_i = \sqrt{I_{4i}}.
\end{equation}
We assumed that an individual fiber begins to bear load only after the fiber-family stretch exceeds a threshold stretch $\lambda_s$. The excess stretch beyond this threshold was represented using a smoothed rectifier function,
\begin{equation}
    \mathcal{R}(\lambda_i;\lambda_s,\tau) = \tau\log\left[1+\exp\left(\frac{1}{\tau}\left(\frac{\lambda_i}{\lambda_s}-1\right)\right) \right],
    \label{eq:smooth_recruitment}
\end{equation}
where $\tau$ controls the sharpness of the transition. Smaller values of $\tau$ approximate a sharper recruitment threshold, whereas larger values produce a smoother transition over a broader stretch range.

Because collagen fibers are expected to have heterogeneous crimping and slack lengths, the threshold stretch $\lambda_s$ was modeled as a distributed variable rather than a single value. The recruitment thresholds were bounded between trainable lower and upper limits, $\lambda_{lb}$ and $\lambda_{ub}$, and described by a probability density function $p(\lambda_s)$. The recruited stretch for the fiber family $i$ was defined as the expectation of the smoothed excess stretch over this distribution,
\begin{equation}
    \lambda_i^{\rm{rec}}
    =1+\int_{\lambda_{lb}}^{\lambda_{ub}}
    \mathcal{R}(\lambda_i;\lambda_s,\tau)
    p(\lambda_s)\,d\lambda_s .
    \label{eq:lambda_rec}
\end{equation}
The corresponding recruited invariant was then
\begin{equation}
    I_{4i}^{\rm{rec}}
    =\left(\lambda_i^{\rm{rec}}\right)^2
    =\left[1+\int_{\lambda_{lb}}^{\lambda_{ub}}
    \mathcal{R}(\lambda_i;\lambda_s,\tau)
    p(\lambda_s)\,d\lambda_s
    \right]^2 .
    \label{eq:I4_recruited}
\end{equation}
The additive constant ensures that $I_{4i}^{\rm{rec}}\rightarrow 1$ when fibers are not recruited, so that the recruited invariant remains consistent with the reference configuration.

Two probability density functions were implemented for the recruitment threshold distribution: a beta distribution and a shifted, truncated lognormal distribution. For the beta distribution, the threshold stretch was rescaled to the unit interval as 
\begin{equation}
    u =
    \frac{\lambda_s-\lambda_{lb}}
    {\lambda_{ub}-\lambda_{lb}}
\end{equation}
on $[0,1]$, with density
\begin{equation}
    p_{\beta}(u)=\frac{u^{\alpha-1}(1-u)^{\beta-1}}{B(\alpha,\beta)},
    \qquad
    B(\alpha,\beta) = \frac{\Gamma(\alpha)\Gamma(\beta)}{\Gamma(\alpha+\beta)} .
    \label{eq:beta_pdf}
\end{equation}
where  $\Gamma(.)$ denotes the Gamma function, and the shape parameters $\alpha$ and $\beta$ control whether recruitment is concentrated near early, intermediate, or late threshold stretches within the admissible interval, without altering the bounds $\lambda_{lb}$ and $\lambda_{ub}$.
For the shifted lognormal distribution, we defined
\begin{equation}
    y = \lambda_s-\lambda_{lb},
    \qquad
    y_{ub} = \lambda_{ub}-\lambda_{lb},
\end{equation}
and used a truncated density over $0<y\leq y_{ub}$,
\begin{equation}
    p_{\log}(y) = \frac{1}  {Z\,y\,\sigma\sqrt{2\pi}} \exp\left[ -\frac{(\ln y-\mu)^2}{2\sigma^2} \right],
    \qquad 0<y\leq y_{ub}.
    \label{eq:lognormal_pdf}
\end{equation}
where $\sigma$ is the scale/shape parameter, and controls the spread and skewness of the recruitment-threshold distribution, $\mu$ is the location parameter and controls where the recruitment threshold is centers, and $Z$ is the  normalization constant, defined as
\begin{equation}
    Z = \Phi\left( \frac{\ln y_{ub}-\mu}{\sigma}\right) = \frac{1}{2}\left[1+\operatorname{erf} \left(\frac{\ln y_{ub}-\mu}
    {\sigma\sqrt{2}}\right)\right].
    \label{eq:lognormal_Z}
\end{equation}
The recruitment integral was evaluated numerically using a 12-point Gauss--Legendre quadrature over the normalized interval $[0,1]$. Recruitment parameters, including $\lambda_{lb}$, $\lambda_{ub}$, distribution shape parameters, and $\tau$, were learned during CANN training (see Supplementary Section \ref{loss}).

\subsection{Model selection}
To obtain a compact strain-energy density function, we pooled the retained terms identified from the subject-specific CANN fits and applied a four-stage model selection workflow: 1) sensitivity analysis to rank influential terms, 2) correlation-based pruning to remove pairwise redundant terms, 3) within-family collapse to simplify the constitutive representation, and 4) refitting and evaluation of the resulting compact candidate models.

\subsubsection{Sensitivity analysis} 
The goal of these step was to evaluate the importance of the pooled terms for the overall strain-energy response across subjects. To assess robustness across deformation ranges, the sensitivity evaluation was performed over two domains: physiological and extended. Basis function were normalized, and coefficients sampling ranges were based on the variability across subject-specific CANN fits. The total-order Sobol indices were used to quantify and rank the influence of the term, and the final sensitivity rank was defined as the mean rank across the both domains. Additional details are provided in Supplementary Section ~\ref{sup_sobol-selection}.

\subsubsection{Correlation-based pruning}
Because different terms can produce the highly similar mechanical response across deformation domains, we performed a correlation-based pruning to identify redundant terms while preferentially retaining those with higher Sobol rankings. A term was retained only when its absolute Pearson correlation with every previously retained term, averaged across the the physiological and extended deformation domains, was below a prescribed threshold. This procedure reduced redundancy while preserving the most influential terms (see Supplementary Section ~\ref{sup_methods:cor-pruning}).

\subsubsection{Within-family reduction}
Because correlation-based pruning was performed pairwise, multiple terms associated with the same invariant or strain measure could remain after pruning. Within-family reduction was therefore used to replace these multiple terms with a single representative function. For each family represented by two or more terms, candidate representatives were drawn from the original CANN function library and evaluated for their ability to reproduce the aggregate response of that family ( see Supplementary Section ~\ref{sup_methods:family-collapse}).

\subsubsection{Candidate refitting}
When within-family reduction identified more than one plausible representative for a constitutive family, the alternative combinations were used to construct a set of candidate reduced SEFs. Each candidate was independently refitted to the subject-specific biaxial data using nonlinear parameter vectors generated by Latin hypercube sampling within bounds informed by the subject-specific CANN fits. For each nonlinear parameter vector, the linear coefficients were determined by nonnegative least squares, and the solution with the lowest stress-fitting loss was used to initialize a bounded local refinement. The resulting candidate models were then compared based on their \(R^2\) values and against the corresponding model prior to within-family reduction. Details and parameter choices can be found in Supplementary Section ~\ref{supp_methods_parameter_search_reduced}.

\subsection{Ablation analysis}
\label{methods_ablation}
Ablation analysis was used to evaluate the contribution of individual terms and modeling choices in the final reduced model. Selected terms or groups of terms were removed, and each ablated model was independently refit to each sample's biaxial data using the same parameter bounds, sampling strategy, optimization settings, and loss function as the full model. Performance was quantified as the percent change in RMSE relative to the full model,

\[
\Delta\mathrm{RMSE} =
\frac{\mathrm{RMSE}_{\mathrm{ablated}}-\mathrm{RMSE}_{\mathrm{full}}}
{\mathrm{RMSE}_{\mathrm{full}}}\times100\%,
\]
with positive values indicating worse fitting performance than the full model.

The purpose of this analysis was diagnostic rather than to further reduce the model. The preceding model-selection framework was used to define the compact constitutive formulation, whereas the ablation analysis was used only to quantify the in-sample contribution of individual terms and modeling choices to the fitted biaxial response.

\subsection{Comparison of reduced SEF with established models} \label{model_evaluation}
The performance of the reduced SEF was evaluated on all nine animal samples using global goodness-of-fit measures and information criteria and compared with three established constitutive formulations previously applied to veins by Sokolis \cite{sokolis_experimental_2013}: the quadratic--Fung model, the quadratic four-fiber-family model, and the neo-Hookean four-fiber-family model. 

The quadratic--Fung model was defined as
\begin{equation}
W_{\mathrm{QF}}=q+K\left[\exp(Q)-1\right],
\label{eq:sokolis_qf}
\end{equation}
where $q = b_{\theta\theta}E_{\theta}^{2} + b_{zz}E_z^{2} + b_{\theta z}E_{\theta}E_z$, and  $Q = c_{\theta\theta}E_{\theta}^{2} + c_{zz}E_z^{2} + c_{\theta z}E_{\theta}E_z$. Here, $E_{\theta}$ and $E_z$ are the circumferential and axial Green--Lagrange strains. The parameters $b_{\theta\theta}$, $b_{zz}$, and $b_{\theta z}$ characterize the low-strain response, whereas $K$, $c_{\theta\theta}$, $c_{zz}$, and $c_{\theta z}$ govern the nonlinear stiffening response.

The quadratic four-fiber-family model was defined as
\begin{equation}
W_{\mathrm{Q4F}} = q + \sum_{m=1}^{4} \frac{k_1^{(m)}}{4k_2^{(m)}} \left[ \exp \left( k_2^{(m)} \left[ \left(\lambda^{(m)}\right)^2-1 \right]^2 \right) -1 \right],
\label{eq:sokolis_q4fiber}
\end{equation}
where $q$ is defined as before, and $k_1^{(m)}$ and $k_2^{(m)}$ control the magnitude and nonlinearity of the response of fiber family $m$, respectively. The stretch of each fiber family was defined as $\lambda^{(m)}
=\sqrt{\mathbf{n}^{(m)}\cdot\mathbf{C}\mathbf{n}^{(m)}}$, where $\mathbf{C}$ is the right Cauchy--Green deformation tensor and $\mathbf{n}^{(m)}$ is the reference direction of fiber family $m$. The first two families represented symmetrically arranged diagonal fibers, and were assigned common material parameters, $k_1^{(1)}=k_1^{(2)},
k_2^{(1)}=k_2^{(2)}$.
The third and fourth families were oriented in the circumferential and axial directions, respectively.

The neo-Hookean four-fiber-family model \cite{Baek2007-dw} was defined using the same four fiber directions, but replaced the quadratic low-strain contribution with an isotropic neo-Hookean term,
\begin{equation}
W_{\mathrm{NH4F}}
=\mu(I_1-3)+\sum_{m=1}^{4}\frac{k_1^{(m)}}{4k_2^{(m)}}\left[\exp\left(k_2^{(m)}
\left[\left(\lambda^{(m)}\right)^2-1\right]^2\right)-1\right],
\label{eq:sokolis_nh4fiber}
\end{equation}
where $\mu$ is a stress-like material parameter and $I_1=\operatorname{tr}\mathbf{C}$ is the first invariant of the right Cauchy--Green tensor.

The quadratic--Fung model provides a fully phenomenological description of the orthotropic response, whereas the two four-fiber formulations are motivated by structural information about diagonal, circumferential, and axial fiber families. The difference between the two four-fiber models lies in the representation of the low-strain response: the quadratic model permits direction-dependent low-strain stiffness, whereas the neo-Hookean model assumes an isotropic matrix contribution. 

The coefficient of determination, $R^2$, was first calculated separately for the circumferential and axial stress responses to provide direction-specific information.  Overall values of $R^2$ were additionally calculated after pooling the circumferential and axial stress observations. These pooled metrics were used as summary measures, whereas the directional values were retained to identify differences in circumferential and axial fitting performance.

The Akaike information criterion, AIC \cite{Akaike1974-co_AIC}, and Bayesian information criterion, BIC \cite{BIC_c4048c8f-6ca9-3965-96a3-653ab8996955}, were used to compare models while penalizing differences in the number of free parameters, here given by linear coefficients and nonlinear recruitment, dispersion, and orientation parameters combined. AIC and BIC were calculated for each specimen individually, using the total number of combined circumferential and axial stress observations fit for that specimen; lower values indicate a more favorable trade-off between fitting accuracy and complexity.

We fitted the literature models independently to every specimen using Latin hypercube sampling with $N=2000$ candidate parameter sets and five independent restarts, followed by bounded L-BFGS-B and Powell refinement.

\section{Results} \label{results}
\subsection{CANN can discover SEFs to fit vein data.}\label{results_cann}
\begin{figure}[!htbp]
    \centering
    \includegraphics[width=1\linewidth]{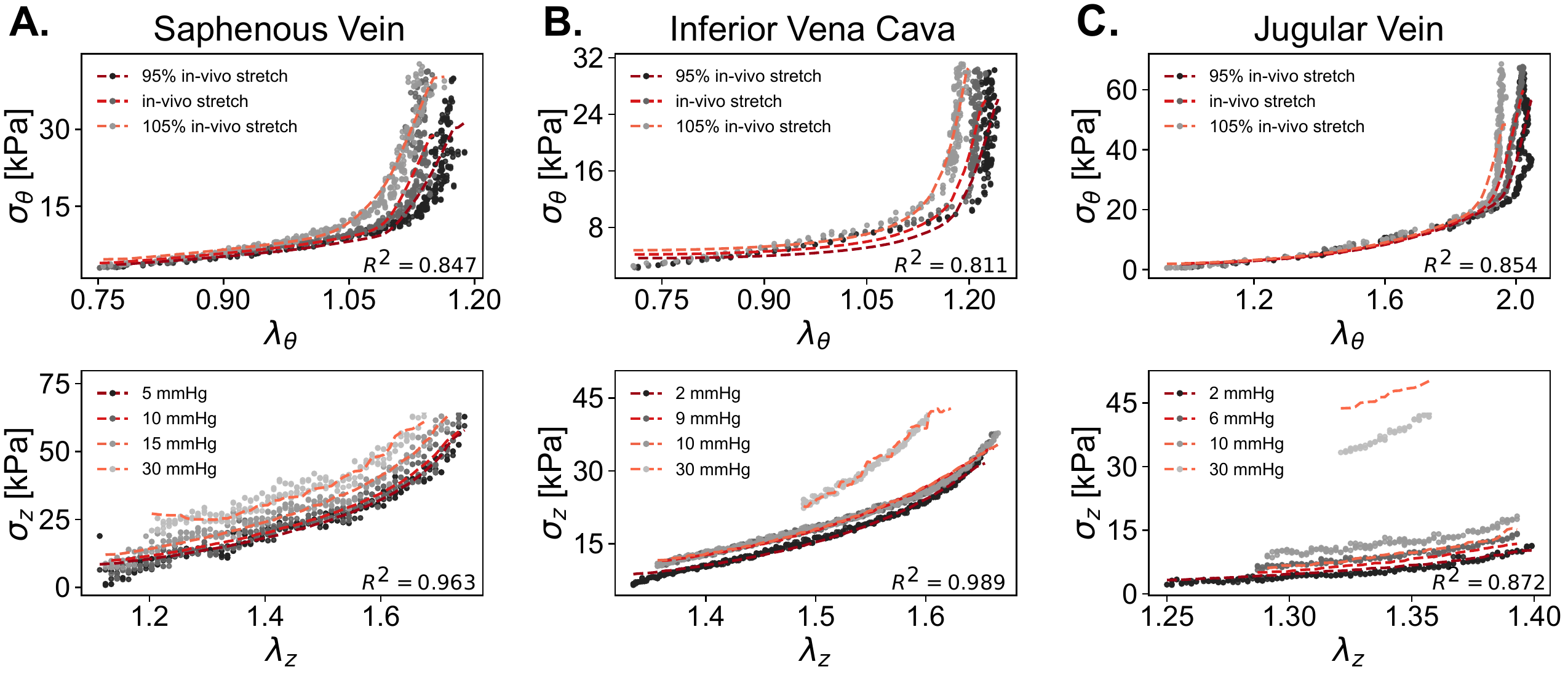}
    \caption{
Representative subject-specific CANN fits to biaxial mechanical data from the three anatomical groups. 
\textbf{(A)} Saphenous vein, \textbf{(B)} inferior vena cava, and \textbf{(C)} jugular vein. 
Top panels show circumferential Cauchy stress, $\sigma_\theta$, as a function of circumferential stretch, $\lambda_\theta$, at three imposed axial stretches. Bottom panels show axial Cauchy stress, $\sigma_z$, as a function of axial stretch, $\lambda_z$, at the indicated luminal pressures. Experimental measurements are shown by closed circles, and CANN predictions are shown by dashed curves. Note the different ranges for stress and stretch across the three different types of veins. The fits for the remaining six samples are shown in Supplemental Figure \ref{supp:remaining_discovered_models}}.
    \label{fig:sample_CANN_fit}
\end{figure}

We trained our CANN architecture separately on the biaxial datasets from each of the nine specimens listed in Table~\ref{tab:training_points_by_vein} and found that it fit the biaxial mechanical behavior well across the saphenous vein, inferior vena cava, and jugular vein samples, despite substantial differences in deformation range, stress magnitude, and degree of nonlinearity. Representative fits are shown in Fig. \ref{fig:sample_CANN_fit}. In the circumferential direction, the models captured the compliant low-stretch region followed by the rapid stiffening at higher stretches, as well as the shift in response produced by changes in axial stretch. In the axial direction, the CANNs reproduced the pressure-dependent separation of the force--length curves across the tested loading protocols. Fits for the remaining specimens are provided in the Supplementary Materials. 

\begin{figure}[!htbp]
    \centering
    \includegraphics[width=1\linewidth]{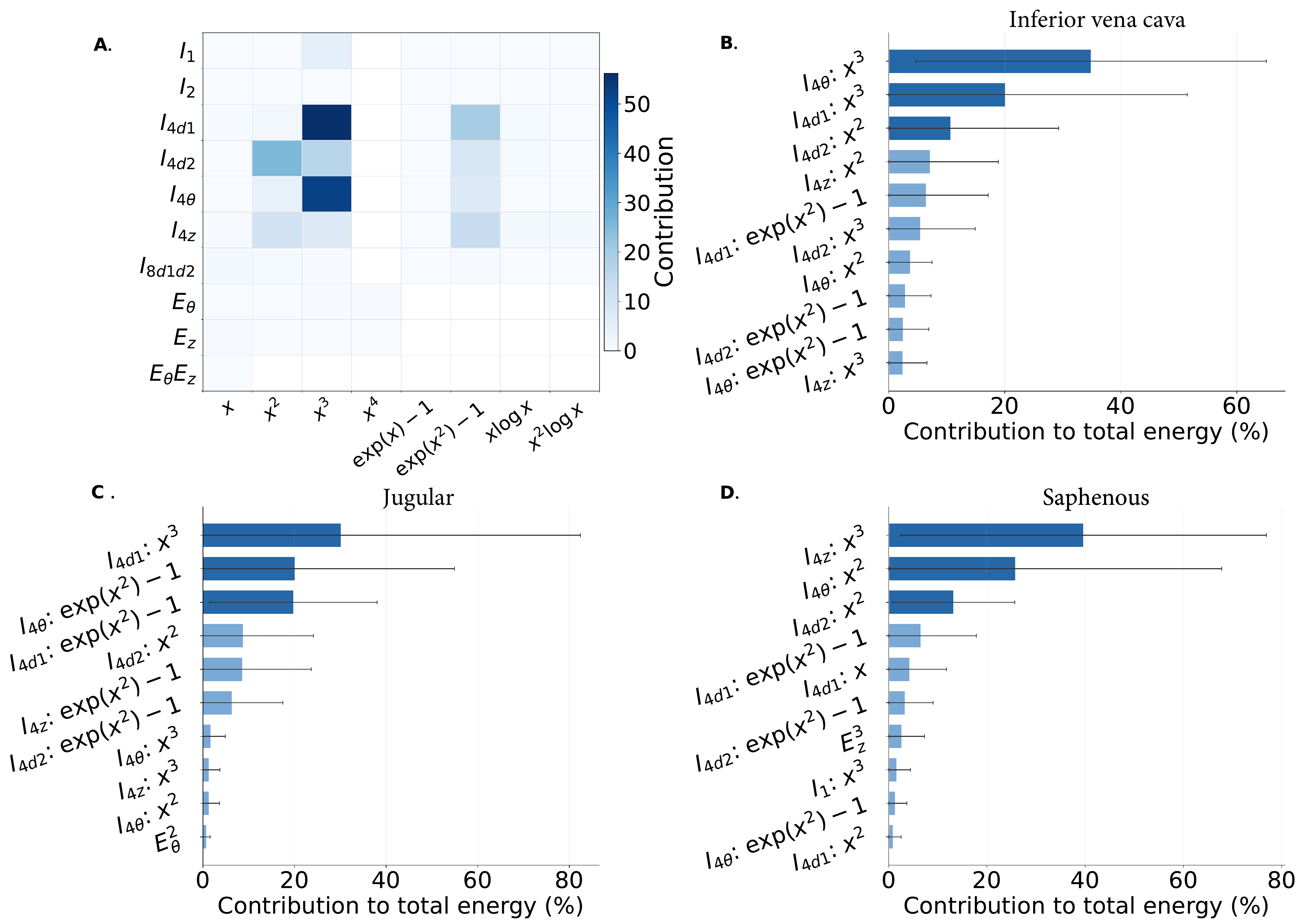}
    \caption{\textbf{Relative contributions of the surviving CANN terms to the total strain-energy density.} (A) Heat map summarizing the contribution of each retained invariant basis-function pair across the venous datasets. (B-D) Ranked contributions for the inferior vena cava, jugular, and saphenous veins, respectively; bars show the mean contribution to total energy, with error bars indicating inter-sample variability. Across tissues, the response is primarily governed by anisotropic $I_4$-based terms, whereas isotropic and $E$-based terms contribute less.}
    \label{fig:cann_surviving_terms}
\end{figure}

Due to intrinsic differences and inter-sample variability, each sample's discovered SEF retained a different subset of terms (Fig.~\ref{fig:cann_surviving_terms} A). The learned strain-energy responses were dominated primarily by anisotropic $I_4$-based terms with fiber recruitment, but the specific invariants, basis functions, and relative contributions differed for the inferior vena cava, jugular, and saphenous veins  (Fig.~\ref{fig:cann_surviving_terms} B-D). Isotropic and strain-based terms generally contributed less to the total energy. 

\subsection{Sensitivity analysis identified recurring SEF terms between samples} \label{results_sobol}
Following specimen-specific CANN training, we pooled the retained terms from the discovered SEFs across all nine specimens into a common 40-term candidate library and performed Sobol sensitivity analysis. Several terms within this pooled library showed high total-order sensitivity, although their relative rankings differed between the physiological and extended deformation domains, indicating that the apparent influence of individual candidate terms depended on the range of deformation over which the strain-energy response was evaluated.

On the physiological domain, the highest-ranked candidates were dominated by functions of the isotropic invariants $I_1$ and $I_2$, together with selected $I_{4d_1}$ contributions. In contrast, the extended domain ranking emphasized the diagonal fiber families $I_{4d_1}$ and $I_{4d_2}$ and the quadratic circumferential-strain term $E_\theta^2$. Thus, the sensitivity ranking reflected both the constitutive family and the deformation range over which the strain-energy response was evaluated (see Fig. \ref{fig:selection_results}A).

Examination of the complete candidate library showed that several polynomial, exponential, and logarithmic functions of the same invariant often received similar total-order Sobol indices (Supplementary Fig.~\ref{fig:supp_sobol_all_terms}). Thus, highly ranked terms associated with the same invariant could represent alternative mathematical descriptions of a similar constitutive response rather than independent mechanical contributions. The Sobol analysis therefore supported the importance of particular constitutive families more strongly than the uniqueness of individual basis functions within those families.  Moreover, because the coefficient sampling ranges were derived from the cross-animal variability of the trained coefficients, the resulting sensitivity indices depended on the variability observed in this cohort. They should therefore be interpreted as cohort-informed measures of sensitivity rather than as universal measures of the mechanical importance of each term.

\subsection{Correlation-based pruning balanced model compactness and fitting performance}\label{results_decollinearization}
We sequentially filtered the retained terms according to the similarity of their evaluated strain-energy responses over the physiological and extended deformation grids, as described in ~\ref{sup_methods:cor-pruning}. 

A larger value of $\rho_{\mathrm{crit}}$ makes the pruning less restrictive, allowing more terms to coexist. The retained terms contained 8, 8, 8, 12, 13, 16, and 24 terms at $\rho_{\mathrm{crit}}=0.90$, $0.93$, $0.95$, $0.96$, $0.97$, $0.98$, and $0.99$, respectively. The corresponding cohort-averaged $R_w^2$ values are as shown in Fig.~\ref{fig:selection_results}B and supplementary table~\ref{tab:supp-rho-performance}. 

 The largest improvement occurred between $0.96$ to $0.97$, where adding only one term increased the $\overline{R_w^2}$ from 0.883 to 0.908. This improvement was driven primarily by the inferior vena cava specimens, for which the $R^2$ increased from $0.8575$ to $0.9303$, corresponding to a relative improvement of $8.49\%$. The jugular and saphenous groups exhibited substantially smaller threshold-dependent changes.(Fig.~\ref{fig:selection_results}B--C).

Despite changes in the specific nonlinear function associated to each invariant and strain, the principal constitutive families were stable across  $\rho_{\mathrm{crit}}=0.96$, $0.97$, and $0.98$ (Table~\ref{tab:rho-selected-terms}). All three pruning thresholds contained contributions associated with $I_1$, $I_2$, the axial and circumferential fiber families, diagonal-fiber interaction through $I_{8d1d2}$, and circumferential--axial coupling through $E_\theta E_z$. The principal differences occurred  within the diagonal-fiber and strain-based contributions. 

At $\rho_{\mathrm{crit}}=0.96$, the 12-term basis contained one diagonal family, together with separate fourth-order circumferential and axial strain terms. At $\rho_{\mathrm{crit}}=0.97$, the second diagonal family was recovered, the independent $E_\theta$ and $E_z$ terms were removed, and additional linear and exponential fiber forms were retained. Increasing the threshold to $\rho_{\mathrm{crit}}=0.98$ expanded the basis from 13 to 16 terms, primarily by introducing additional nonlinear alternatives within families already represented at $\rho_{\mathrm{crit}}=0.97$. These included a quadratic $I_1$ term, a logarithmic form for $I_{4d1}$ and $I_{4z}$, and an alternative exponential term for $I_{4d2}$. 

Although $\rho_{\mathrm{crit}}=0.97$ provided the largest improvement in performance relative to the number of added terms, $\rho_{\mathrm{crit}}=0.98$ was retained as a more inclusive parent basis for the subsequent response-based family collapse.

\begin{table}[t]
\centering
\caption{
Basis functions retained after correlation-based pruning at three correlation thresholds.
Within each row, $x$ denotes the shifted input associated with the corresponding
invariant or strain family.
}
\label{tab:rho-selected-terms}

\small
\setlength{\tabcolsep}{7pt}
\renewcommand{\arraystretch}{1.25}

\begin{tabularx}{\textwidth}{
@{}
>{\raggedright\arraybackslash}p{0.15\textwidth}
>{\raggedright\arraybackslash}X
>{\raggedright\arraybackslash}X
>{\raggedright\arraybackslash}X
@{}
}
\toprule
\textbf{Basis family}
&\textbf{$\rho_{\mathrm{crit}}=0.96$}&\textbf{$\rho_{\mathrm{crit}}=0.97$}&\textbf{$\rho_{\mathrm{crit}}=0.98$}
\\
\midrule
$I_1$&$\exp(x^2)-1$&$\exp(x^2)-1$&\makecell[l]{$\exp(x^2)-1$ \\ $x^2$}
\\[2pt]
$I_2$&\makecell[l]{$x$ \\ $x^2$}&\makecell[l]{$x$ \\ $x^2$}&\makecell[l]{$x$ \\ $x^2$}
\\
\addlinespace[3pt]
$I_{4d2}$&--&$x\log(x+1)$&$\exp(x^2)-1$
\\[2pt]
$I_{4d1}$&\makecell[l]{$x$ \\ $x^3$}&\makecell[l]{$x$ \\ $x^3$}&\makecell[l]{$x$ \\ $x^3$ \\ $x\log(x+1)$}
\\[2pt]
$I_{4z}$&$x^2$&\makecell[l]{$x$ \\ $x^2$}&\makecell[l]{$x$ \\ $x^2$ \\ $x\log(x+1)$}
\\[2pt]
$I_{4\theta}$&$\exp(x^2)-1$&\makecell[l]{$\exp(x)-1$ \\ $\exp(x^2)-1$}&\makecell[l]{$\exp(x)-1$ \\ $\exp(x^2)-1$}
\\[2pt]
$I_{8d1d2}$&\makecell[l]{$x$ \\ $x^2$}&\makecell[l]{$x$ \\ $x^2$}& \makecell[l]{$x$ \\ $x^2$}
\\
\addlinespace[3pt] $E_\theta E_z$ & $E_\theta E_z$ & $E_\theta E_z$   & $E_\theta E_z$
\\[2pt]
$E_\theta$ &$E_\theta^4$&--&--
\\[2pt]
$E_z$&$E_z^4$&--&--
\\
\midrule
\textbf{Total terms}&\textbf{12}&\textbf{13}&\textbf{16}
\\
\bottomrule
\end{tabularx}
\end{table}
\subsection{Response-based family collapse reduced the SEF without loss of performance}
\label{results_family_collapse}

The 16 terms retained at $\rho_{\mathrm{crit}}=0.98$, hereafter denoted CT98, contained multiple mathematical functions associated with the same constitutive family, where each family was defined by its underlying invariant or strain measure. We therefore used response-based family collapse to replace the multiple terms within each family with a single representative function while preserving the energy response of the original CT98 basis.

\begin{figure}[!htbp]
    \centering
    \includegraphics[width=1.0\linewidth]{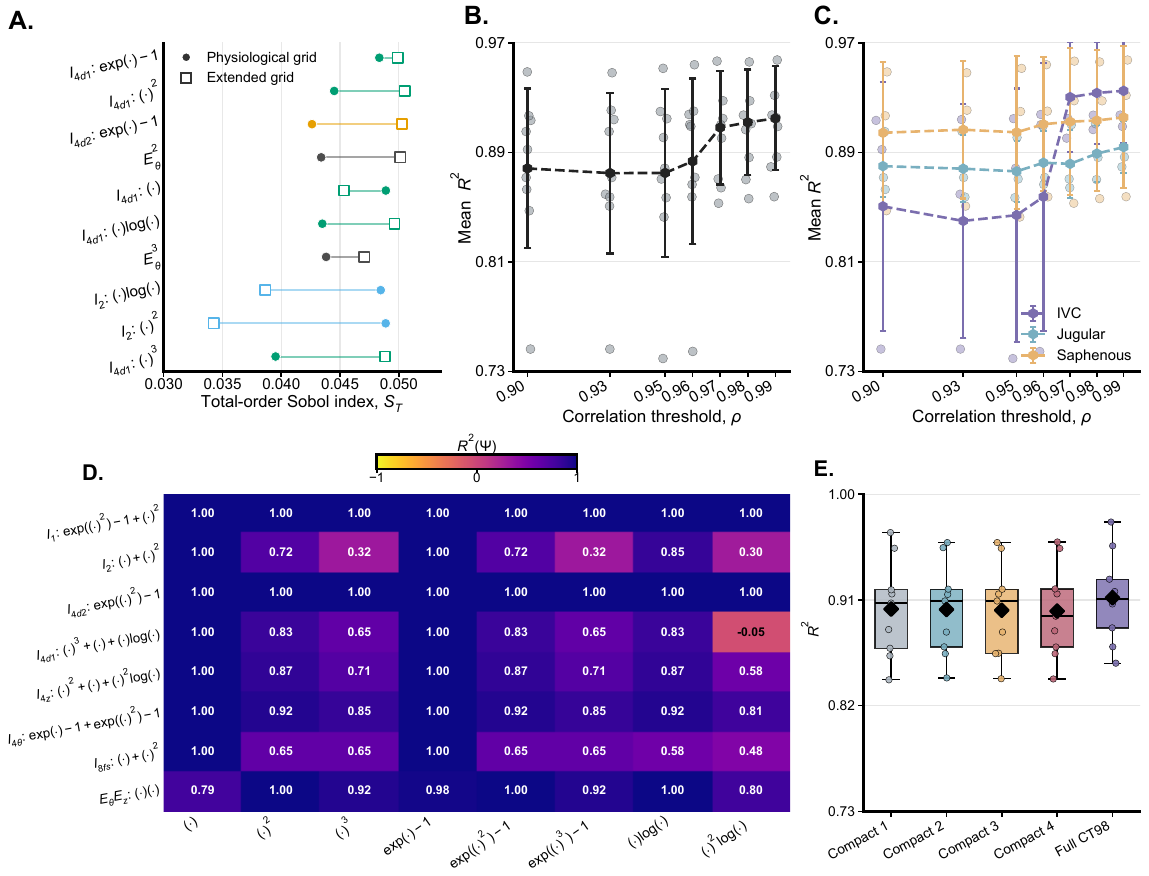}
\caption{
\textbf{Sequential sensitivity ranking, correlation-based pruning, and within-family reduction of the candidate constitutive basis}.
\textbf{(A)} Total-order Sobol indices, $S_T$, for the ten highest-ranked candidate terms evaluated over the physiological and extended biaxial deformation grids. Several alternative functions associated with the same invariant families ranked highly, motivating the subsequent redundancy analysis.
\textbf{(B)} Cohort-averaged $R^2$ across correlation thresholds, $\rho$. Gray markers represent individual subjects, whereas black markers and error bars denote the cohort mean and standard deviation. Performance changed little for $\rho=0.90$--$0.96$ but increased for $\rho\geq0.97$ as additional basis terms were retained.
\textbf{(C)} $R^2$ stratified by vein type. The inferior vena cava showed the greatest threshold-dependent improvement, whereas performance for the jugular and saphenous veins remained comparatively stable.
\textbf{(D)} Pairwise $R^2(\Psi)$ between each aggregated basis-family response and candidate single-term representations. The high agreement observed for several candidates indicates that the aggregated energy responses can be approximated by simpler collapsed terms over the physiological deformation grid.
\textbf{(E)} Distribution of subject-specific $R^2$ for four compact formulations containing 8 terms each and the 16-term model retained at $\rho=0.98$ (Full CT98). The compact formulations loss very little performance while reducing the number of retained terms by half.} 
\label{fig:selection_results}
\end{figure}

For each constitutive family, the contributions of all corresponding CT98 terms were first summed, using their cohort-mean effective coefficients, to define an aggregate family-level energy response. Individual candidate functions from the original constitutive library were then fitted to this aggregate response over the
physiological deformation grid. Candidates satisfying $R_W^2 \geq 0.80$ were retained as plausible single-term representatives of that family. As shown in Fig.~\ref{fig:selection_results}D, several alternative descriptors reproduced the corresponding aggregate family responses with high $R^2(\Psi)$, indicating that multiple functional forms can describe nearly equivalent energy responses over the physiological range.

Agreement with an isolated family response did not necessarily guarantee accurate stress predictions when the representative functions were combined into the complete constitutive equation. The retained family-level candidates were therefore evaluated through a constrained combinatorial search. The $I_1$ and $I_2$ contributions were fixed as linear terms, and the strain-based contribution was fixed as $E_\theta E_z$. The two diagonal families, $I_{4d1}$ and $I_{4d2}$, were constrained to use the same class of constitutive functions and the same coefficients and nonlinear parameters to avoid the development of a net torque due to asymmetry \cite{Humphrey_torque_Emuna2026-zn}. The remaining choices consisted of four shared $I_{4d1}/I_{4d2}$ functions, eight $I_{4\theta}$ functions, five $I_{4z}$ functions, and two $I_{8d1d2}$ functions. Their combinations produced $4\times8\times5\times2=320$ candidate eight-term equations.

Each of the 320 equations was independently fitted to biaxial stress data from each specimen. The nonlinear recruitment, dispersion, and fiber orientations parameters were initially explored using a 256-point Latin hypercube sample. For each sampled nonlinear parameter set, the linear constitutive coefficients were determined by non-negative least squares. The best-performing Latin-hypercube candidates were subsequently refined using L-BFGS-B optimization. 

All 320 equations produced successful fits for all nine specimens, yielding 2880 successful subject-specific refits. Across the complete candidate set, the cohort-averaged coefficient of determination ranged from $\overline{R_w^2}=0.458$ to $0.902$, while the mean combined RMSE ranged from $9.42$ to $4.06$~kPa.

The four highest-performing compact formulations are compared with the original 16-term CT98 model in Fig.~\ref{fig:selection_results}E. Despite reducing the number of retained terms from 16 to eight, all four compact models preserved nearly the same cohort-average performance as CT98. Compact~2 was selected as the reduced constitutive model here because its mean $R^2$ was closest to that of CT98 ($0.902$ versus $0.911$) while exhibiting a comparatively narrow inter-sample spread and stronger performance among the least accurately fitted samples. 

The uniqueness of the selected functional form was subsequently assessed relative to Compact~2. A candidate equation was classified as uniformly near-equivalent when its combined RMSE remained within 5\% of that of Compact~2 for every individual specimen. Sixteen of the 320 candidate equations satisfied this criterion. These models therefore preserved fitting performance across all nine specimens rather than only at the cohort-average level. The near-equivalent set contained several different nonlinear representations of the fiber families, indicating that the eight-family constitutive architecture was more consistently identifiable than the specific mathematical function assigned to every family. The 16 near-equivalent equations and their fitting metrics are reported in Supplementary Materials \ref{tab:near-equivalent-equations-fungd}.

\subsection{A compact, mechanically interpretable SEF for veins}
\label{results_final_model}

We identified a consistent form of the venous SEF across the highest-performing candidate models using this basis-reduction framework described in the previous sections. For the diagonal and circumferential fiber families with fiber recruitment, the invariant $I_{4\alpha}^{\mathrm{rec}}$ (for $\alpha \in \{d_1, d_2, \theta\}$) was computed using the recruitment formulation defined in Section~\ref{fiber-recruit}. The reduced SEF for veins is
\begin{equation}
\begin{aligned}
\Psi_{\mathrm{vein}} = {}& c_1(I_1-3)+c_2(I_2-3) + c_{d}\left[\Phi_c\!\left(I_{4d1}^{\mathrm{rec}}-1\right)+
\Phi_c\!\left(I_{4d2}^{\mathrm{rec}}-1\right)\right] \\
&+c_{\theta}\Phi_{c}\!\left(I_{4\theta}^{\mathrm{rec}}-1\right)
+c_z\Phi_{el}\!\left(I_{4z}-1\right)
+c_{d1d2}\Phi_{el}\!\left(I_{8d1d2}-I_{8d1d2}^{0}\right)\\
&+c_{\theta z}E_{\theta}E_z .
\end{aligned}
\label{eq:proposed_venous_SEF}
\end{equation}
where
\begin{align}
\Phi_{\mathrm{c}}(x)
&=x^3,
\\
\Phi_{\mathrm{el}}(x)
&=\exp(x)-1,
\end{align}
denote quadratic, exponential-quadratic, and exponential-linear responses, respectively, and $I_{8d1d2}^{0}$ denotes the reference-state value of the $I_{8d1d2}$ invariant, as determined from the reference orientations of the two diagonal fiber families. All eight constitutive coefficients were constrained to be non-negative during refitting.
The first two terms in Eq.~\eqref{eq:proposed_venous_SEF} describe the effective isotropic response through linear functions of $I_1$ and $I_2$. The diagonal and circumferential fiber families follow recruited cubic responses, whereas the axial family follows an exponential-linear response without an explicit recruitment operator. The final two terms describe coupling between the diagonal fiber families and direct coupling between circumferential and axial deformation, respectively. All eight constitutive coefficients were constrained to be non-negative during refitting. 

The proposed reduced SEF produced the lowest mean combined RMSE among the compact eight-term models, with a mean combined RMSE of $4.09$, $R_\theta^2=0.901$, and $R_z^2=0.958$. In addition to its in-sample fitting performance, the model retained a direct interpretation in terms of isotropic response, direction-specific fiber recruitment and stiffening, diagonal-fiber interaction, and circumferential--axial coupling.

\begin{figure}[!htbp]
    \centering
    \includegraphics[width=1.0\linewidth]{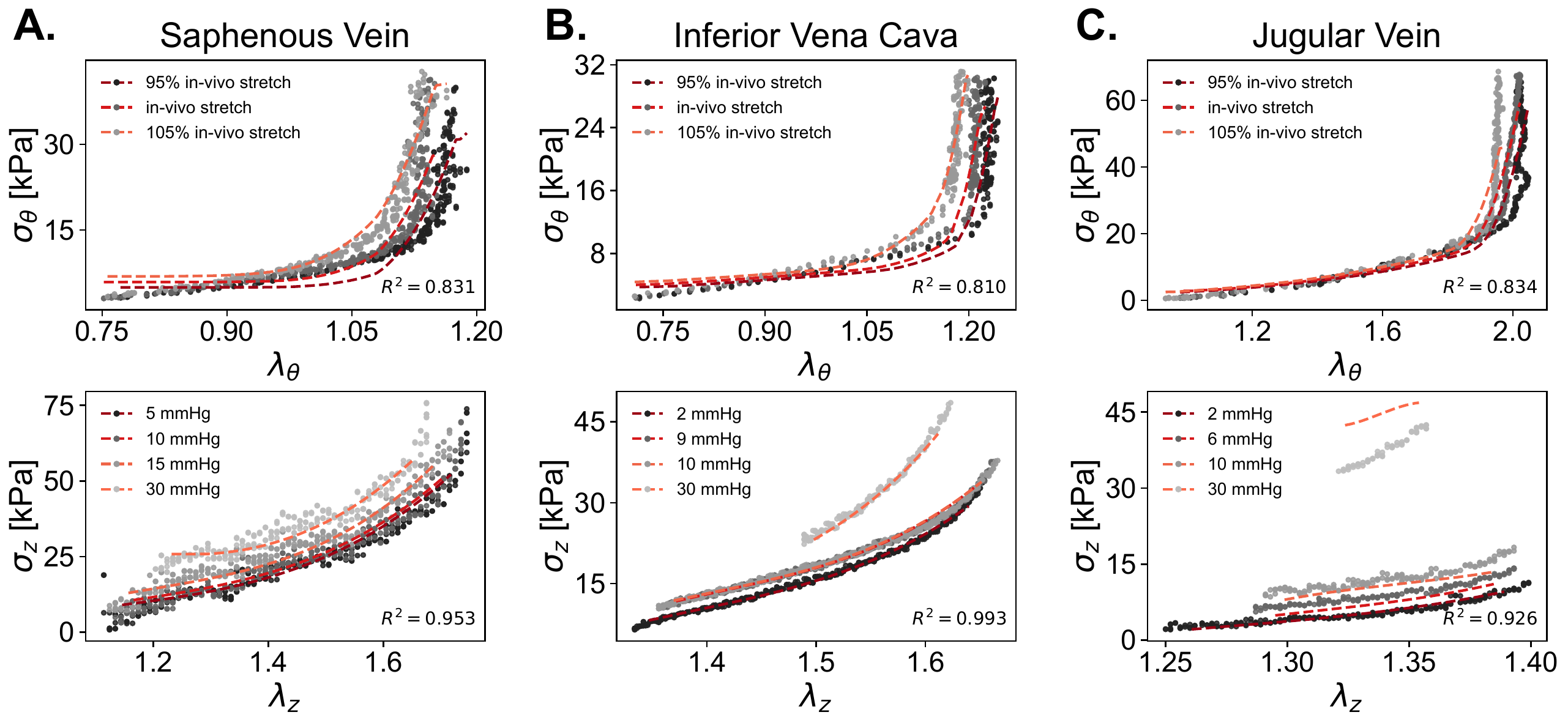}
\caption{
Representative fits of the proposed reduced venous SEF to biaxial mechanical data from the three anatomical groups \textbf{(A--C)}. Top row shows circumferential Cauchy stress, $\sigma_\theta$, as a function of circumferential stretch, $\lambda_\theta$, at 95\%, 100\%, and 105\% of the estimated in vivo axial stretch for representative saphenous, inferior vena cava, and jugular specimens, respectively. 
Bottom row shows axial Cauchy stress, $\sigma_z$, as a function of axial stretch, $\lambda_z$, at the indicated pressure levels for the corresponding specimens. Experimental measurements are shown by circular markers, and predictions from the proposed closed-form SEF are shown by dashed curves. The reported $R^2$ values were calculated by pooling all loading protocols within each panel.}
    \label{fig:proposed_sef_fit}
\end{figure}

Representative fits of the reduced SEF are shown in Fig.~\ref{fig:proposed_sef_fit} for saphenous, inferior vena cava, and jugular vein specimens. The model reproduced the nonlinear circumferential response across the three imposed axial stretches, including the transition from the compliant low-stretch region to rapid stiffening at higher stretches, with pooled circumferential $R^2$ values of $0.831$, $0.810$, and $0.834$ for the representative saphenous, inferior vena cava, and jugular specimens, respectively. The model also captured the pressure-dependent axial response and the separation among the different pressure protocols, with pooled axial $R^2$ values of $0.953$, $0.993$, and $0.926$, respectively.

\subsubsection{Importance of individual terms in the reduced SEF using ablation analysis}
\label{ablation_analysis}
\begin{figure}
    \centering
    \includegraphics[width=1.0\linewidth]{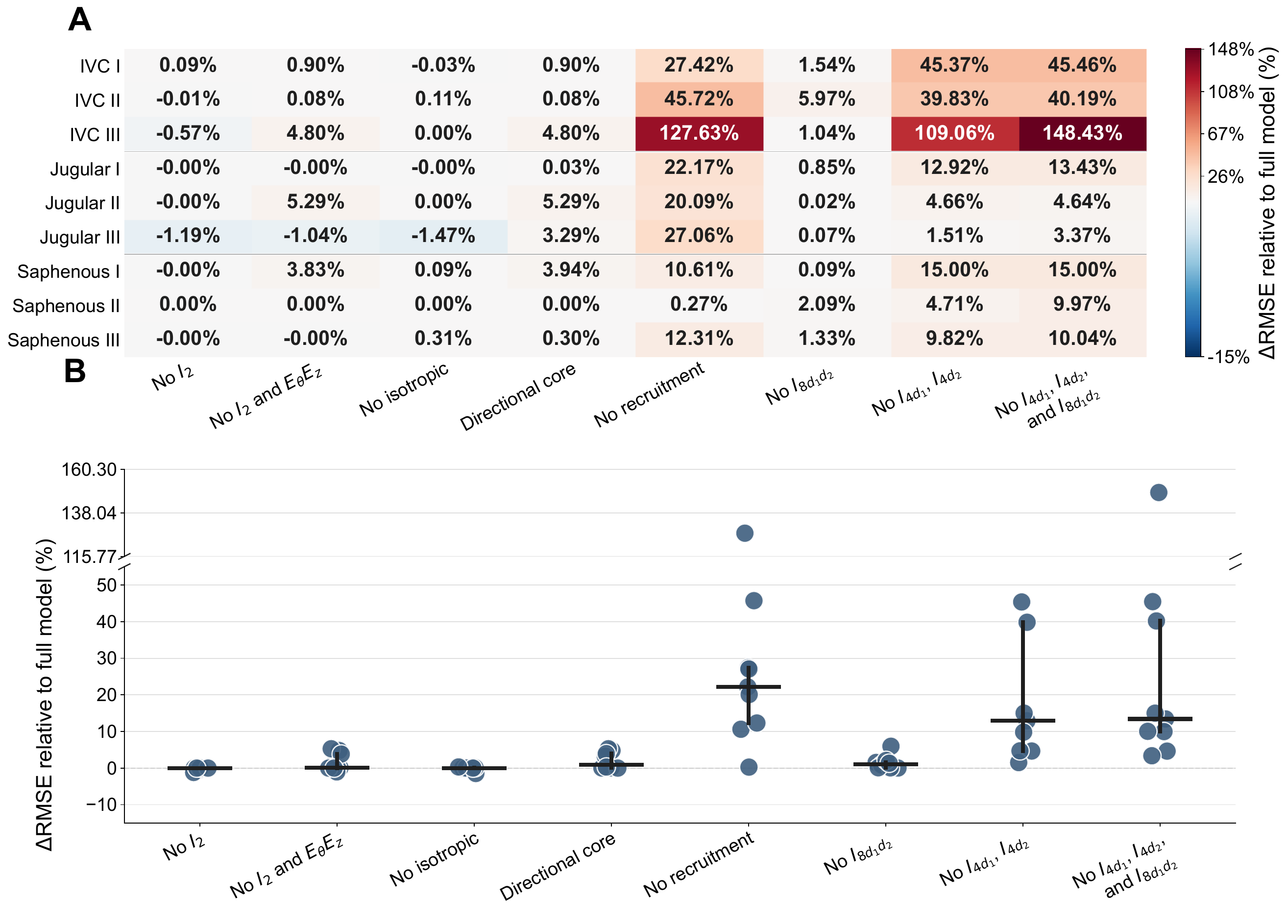}
   \caption{\textbf{Ablation analysis of the proposed SEF.}
(A) Subject-specific percentage change in combined RMSE after removal or modification of selected constitutive terms relative to the full model. Positive values indicate deterioration in fitting accuracy, whereas negative values indicate improvement. Ablation analysis evaluated removal of $I_2$, $I_2$ together with $E_\theta E_z$, both isotropic invariants, all isotropic and direct strain-coupling terms (directional core), removal of recruitment, $I_{8d_1d_2}$, diagonal families, all diagonal contributions and their interaction. (B) Distribution of $\Delta\mathrm{RMSE}$ across samples for each ablation. Points denote individual specimens, and black horizontal markers with vertical bars denote the median and interquartile range. The broken vertical axis displays large errors produced by selected structural ablations.}
    \label{fig:ablation_rmse_summary}
\end{figure}

Ablation analysis was used to determine which components of the reduced SEF were necessary to preserve fitting accuracy. The full model contained four directional terms, \(I_{4d_1}\), \(I_{4d_2}\), \(I_{4\theta}\), and \(I_{4z}\); the diagonal interaction term \(I_{8d_1d_2}\); the isotropic invariants \(I_1\) and \(I_2\); and the direct circumferential--axial strain-coupling term \(E_\theta E_z\). Figure~\ref{fig:ablation_rmse_summary}A shows the specimen-specific change in RMSE relative to the full model, whereas Fig.~\ref{fig:ablation_rmse_summary}B summarizes the distribution across specimen.

We found that isotropic and direct strain-coupling terms could be removed with limited change in accuracy. Removing \(I_2\) changed the RMSE by less than \(-1.2\%\) for every specimen. Removing both isotropic invariants while retaining \(E_\theta E_z\) produced changes between \(-1.04\%\) and \(5.29\%\). The directional-core model, obtained by additionally removing \(E_\theta E_z\) and retaining only \(I_{4d_1}\), \(I_{4d_2}\), \(I_{4\theta}\), \(I_{4z}\), and \(I_{8d_1d_2}\), increased the RMSE to \(3.29\%\) for jugular III and \(3.94\%\) for saphenous I, with no further changes for the remaining specimens.

In contrast, removal of the recruitment operators substantially reduced model performance. As shown in Fig.~\ref{fig:ablation_rmse_summary}, the median RMSE increase was \(22.17\%\). The largest change occurred for IVC III, for which the RMSE increased by \(127.63\%\). Recruitment was therefore required to reproduce the nonlinear response consistently across the cohort.

The diagonal interaction term \(I_{8d_1d_2}\) had a relatively small independent contribution; removing this term alone increased the RMSE by less than \(6\%\) for every specimen. In contrast, removing the diagonal directional terms, showed a stronger and more specimen-dependent effect, increasing RMSE by up to \(109.06\%\) for IVC III. Simultaneous removal of the diagonal interaction term and diagonal directional terms produced an even larger deterioration, yielding RMSE exceeding  \(10\%\) in eight of the nine subjects and reaching \(148.43\%\) for IVC III. 

Taken together, these results indicate that the isotropic invariants and direct circumferential--axial strain-coupling term contributed little to the overall fit, whereas fiber recruitment and the diagonal directional terms were important for capturing the responses of several specimens. Although \(I_{8d_1d_2}\) had a limited independent effect, the pronounced deterioration following removal of the complete diagonal contribution indicates that the diagonal directional and interaction terms collectively provide mechanical behavior that cannot be reproduced by the circumferential and axial terms alone. 

\subsubsection{The reduced SEF has significantly improved performance compared to literature models}
\label{results_comparison}

\begin{figure}[!htbp]
\centering
\includegraphics[width=1.0\textwidth]{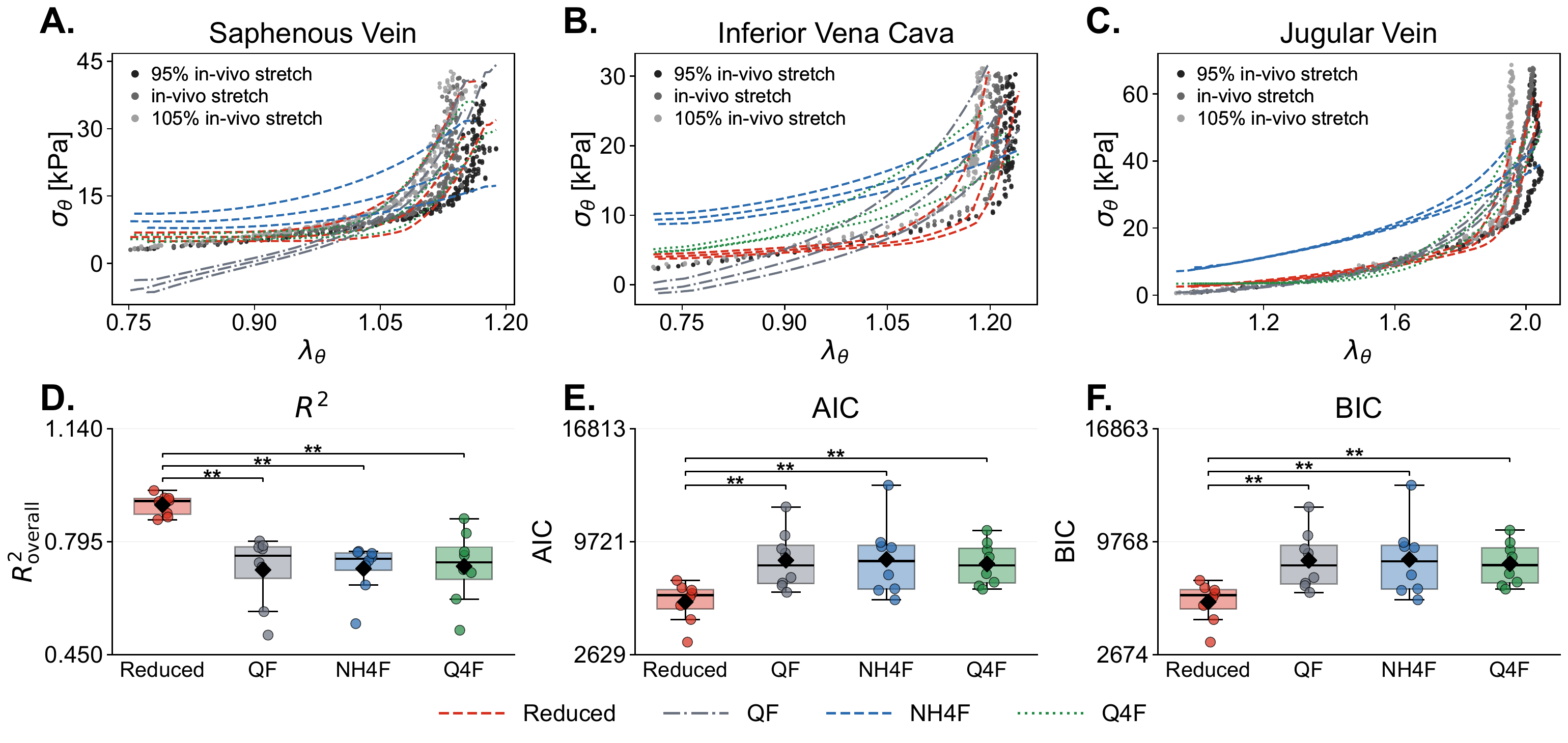}
\caption{
Comparison of the reduced SEF with the quadratic--Fung (QF), quadratic four-fiber-family (Q4F), and neo-Hookean four-fiber-family (NH4F) models across venous samples. (A--C) Experimental circumferential stress--stretch responses and model predictions for representative saphenous, inferior vena cava, and jugular vein samples at 95\%, 100\%, and 105\% of the in vivo axial stretch. Gray symbols denote experimental measurements, while the red dashed, blue dash-dotted, and green dotted curves denote predictions from the proposed, NH4F, and Q4F models, respectively. Box plots show the distributions of (D) overall coefficient of determination, $R^2_{\mathrm{overall}}$, (E) Akaike information criterion, and (F) Bayesian information criterion. Individual circles represent subject-specific values, horizontal lines indicate medians, and black diamonds indicate cohort means. Higher $R^2_{\mathrm{overall}}$ and lower, AIC, and BIC indicate better performance. Statistical annotations denote one-sided paired Wilcoxon signed-rank comparisons between each literature model and the proposed SEF, with significance levels defined as $^{***}p<0.001$, $^{**}p<0.01$, $^{*}p<0.05$, $\cdot\,p<0.10$, and ns, not significant.
}\label{fig:model_performance}
\end{figure}

The reduced SEF achieved consistently higher overall coefficients of determination than the three literature models (Fig.~\ref{fig:model_performance} A). The improvement in $R^2_{\mathrm{overall}}$ was significant relative to each literature formulation based on one-sided paired Wilcoxon signed-rank tests ($p<0.01$).

Additionally, it yielded lower AIC and BIC values than each literature model (Fig.~\ref{fig:model_performance}C--D). These differences were significant for all paired comparisons ($p<0.01$). 

\section{Discussion}\label{discussion}
This work advances venous constitutive modeling in two principal ways. First, it introduced a systematic, data-driven framework in which a CANN, modified to include a broader candidate library, is trained on biaxial mechanical testing data to discover terms needed to reproduce venous mechanical behavior. Second, it applied the framework on three different types of veins (saphenous, inferior vena cava, and jugular) to obtain a reduced constitutive model, achieving a more favorable balance of fitting accuracy and model complexity than the existing literature models considered in the study.

The CANN formulation extended previous constitutive-discovery approaches \cite{linka_constitutive_2021, peirlinck_universal_2024, Linka2023-ka_CANN_constraints}, to include an effective isotropic response, direction-specific fiber contributions, interaction between diagonal fiber families, and circumferential-axial coupling. The network was fitted to each specimen separately rather than pooled data, since a single pooled model would necessarily yield one common set of weights and nonlinear parameters, which would be unable to capture the anatomical differences and inter-sample variability. The goal was thus to identify a common constitutive architecture with specimen-specific parameters, not a single universal parameter set. Following CANN-based discovery, sensitivity ranking, correlation-based pruning, family-level collapse, and specimen-specific refitting, an eight-term formulation (Eq. \ref{eq:proposed_venous_SEF}) preserved the fitting performance of the 16-term parent basis with minimal loss of accuracy. Within the common nine-specimen comparison cohort, the proposed SEF also achieved higher overall fitting accuracy and more favorable information criteria than previously reported models used for arteries and veins, namely quadratic-Fung, quadratic four-fiber-family, and neo-Hookean four-fiber-family models.

The proposed model can be viewed as combining four mechanical contributions: an effective isotropic background response ($I_1$, $I_2$), direction-specific fiber terms ($I_{4\theta}$, $I_{4z}$, $I_{4d1}$, $I_{4d2}$), diagonal-fiber coupling ($I_{8d1d2}$), and a direct circumferential--axial strain-coupling term ($E_\theta E_z$). Unlike the fiber-recruitment and fiber-coupling terms, which suggest identifiable structural features, the strain-coupling term has no microstructural counterpart and is more naturally interpreted as an addition to the effective ground-matrix response rather than a distinct fiber mechanism. Notably, the four fiber families did not share a single functional form, diagonal and circumferential families followed a recruited cubic response, whereas the axial family followed an exponential-linear response. This is a departure from conventional practice, in which all fiber families are typically assigned the same functional template (most often a single exponential form) with only the coefficients allowed to vary. The different forms selected here are consistent with the distinct roles these families play in the compliant-to-stiffened transition zone. For example, the simpler exponential-linear form for the axial family reflects its comparatively modest stiffening relative to the sharply stiffening circumferential response. The retention of the $E_\theta E_z$ coupling term is also consistent with prior venous work that introduced strain-based terms instead of invariant-based descriptors for the non-fiber response \cite{sokolis_experimental_2013}.

The proposed model retains useful features of earlier venous formulations while addressing several of their limitations. The quadratic-Fung model contains the $E_\theta E_z$ coupling term, and the neo-Hookean and quadratic four-fiber models contain exponential $I_4$ invariant terms, yet none of these formulations proved flexible enough to fit the circumferential and axial venous responses well simultaneously \cite{sokolis_experimental_2013}. The proposed model retains both of these features -- direct strain coupling and exponential fiber terms -- while adding progressive fiber recruitment for the diagonal and circumferential fiber families and cross-fiber interaction between the diagonal families. These additions, rather than being prescribed a priori, were selected because the data consistently preferred them over the corresponding literature forms.

The most substantive addition is the recruitment term, motivated by non-venous collagen-rich tissues in which distributions of slackness represent the progressive straightening and load-bearing engagement of undulated fibers \cite{Famaey2015-vk_review_of_bridging_veins, li_multi-scale_2023} — a mechanism well suited to veins, whose circumferential response shows a long compliant region followed by abrupt stiffening. Removing recruitment in our cohort increased the population median RMSE by 22.17\%, and increased RMSE by 127.63\% for one inferior vena cava specimen. Thus, recruitment was the feature with the most consistent empirical support among the tested extensions.  

While this supports the empirical need of recruitment, it does not establish a one-to-one correspondence between the fitted recruitment distribution and collagen crimp. Within the present model, recruitment should be interpreted as an effective constitutive mechanism that captures a distributed directional transition from compliant to stiffer behavior. It provides a constrained means of representing the shape of the venous stiffening transition without requiring a large collection of independent high-order polynomial or exponential terms. Simultaneous imaging of fiber waviness, alignment, or engagement during biaxial loading would be needed to determine whether the fitted recruitment thresholds correspond quantitatively to fiber-level recruitment.

The directional ablations indicated that the diagonal fiber families provided a substantial contribution to the model's fitting accuracy. Removing the diagonal terms increased RMSE by $10\%$  in eight out of nine specimens, with the largest effect observed for IVC III ($109\%$). Simultaneous removal of diagonal terms and their interaction term $I_{8d1d2}$ produced the largest deterioration seen in the entire ablation analysis, $148.43 \%$ for IVC III. This pronounced, specimen-dependent sensitivity indicates that axial and circumferential fiber families alone are not sufficient to capture venous mechanical behavior, and that the diagonal families provide a mechanically distinct contribution that cannot be recovered by the orthogonal axial-circumferential pair alone.

We also observe that when the CANN was trained on data from nine individual animals, it fit the measured biaxial data well but did not select the same individual basis functions across specimens. This is expected when the available experiments sample only a finite deformation domain: several nonlinear functions can generate closely similar energy and stress responses over such a domain. It is also expected given the substantial variability in mechanical response across samples, which alone would drive different basis-function selections. The subsequent reduction procedure was therefore designed to separate recurring response families from alternative mathematical descriptions of the same response. The fact that 16 of 320 compact equations remained within 5\% of the proposed model's combined RMSE for every specimen makes this point explicit. The present data identify a robust family-level architecture more strongly than they identify a unique nonlinear function for every fiber family. Accordingly, the proposed constitutive relation should be viewed as a parsimonious representative of a near-equivalent model class, not as definitive evidence that its exact cubic and exponential forms are uniquely determined by venous microstructure.

The results provide more qualified support for the non-fiber terms. Removing the isotropic contributions had little effect for most specimens, whereas direct circumferential–axial strain coupling produced a localized improvement, most notably in selected jugular and vena cava samples. Thus, the data do not require every retained term to be equally important in every vein. Instead, the full equation provides a common form that can accommodate inter-specimen variation while permitting some coefficients to become small when a contribution is not needed. This distinction matters for model interpretation: inclusion of $E_{\theta}E_z$ and the diagonal interaction invariant demonstrates that coupled deformation responses improve the model class, but it does not establish that each coupling contribution is independently identifiable for every specimen. Whether a more compact form would sufficiently describe vein mechanics would require richer loading paths and deliberately designed protocols that vary axial stretch and pressure independently over a wider range and on a larger dataset.

A related question concerns the mathematical structure of the resulting constitutive model. Polyconvexity is often a property of interest; along with coercivity, it can guarantee existence of minimizers. On its own, it guarantees ellipticity, which in turn is desirable for finite element simulations \cite{polyconvexity_Klein2026-ja}. Our CANN-network and the proposed reduced model do not guarantee polyconvexity due to the $E_\theta E_z$ coupling term and the $I_8$ cross-fiber terms, and do not satisfy the strong ellipticity condition or the Legendre-Hadamard inequality \cite{ellipticity_Wilber2002-wt} for all sets of parameters. Several of the models in the literature that we considered for veins do not possess these properties either. The absence of these properties does not necessarily imply convergence problems in practice, but it identifies an area warranting further investigation.

Several limitations define the appropriate scope of the conclusions. First, the cohort contained nine healthy ovine specimens from three anatomical locations. The shared constitutive architecture across these samples is encouraging, but it does not establish generalization across species, sex, age, disease state, or human vein-graft conditions. Second, stresses and stretches were inferred under radially averaged, axisymmetric, incompressible deformations. Residual stress, wall-layer heterogeneity, spatially varying fiber architecture, and other complexities may therefore be absorbed into effective model parameters. Third, the candidate library constrained what the CANN could discover, and the Sobol ranking depended on normalized basis columns and cohort-informed coefficient perturbation ranges. It ranks terms for this discovery setting; it is not a universal measure of mechanical importance. Finally, the model-selection and performance results were evaluated using the same nine specimens and biaxial loading data used for model discovery and parameter fitting, rather than on an independent validation dataset. The reported accuracy should therefore be described as evidence of compact descriptive adequacy, not as an estimate of out-of-sample predictive accuracy.

Future studies should examine whether the proposed architecture predicts responses that were not used during fitting. Useful validation experiments would include independent cohorts from additional venous locations. A hierarchical implementation could then estimate a population-level architecture while allowing specimen- and location-specific parameter variation. Pairing these experiments with histology or multiphoton imaging would also test whether the fitted recruitment and diagonal-family parameters track collagen crimp, orientation dispersion, and remodeling. If validated in this way, the model could provide a practical constitutive relation for graft, fistula, and tissue-engineered vascular-graft simulations in which venous compliance and the onset of circumferential stiffening are mechanically consequential. 

\section{Conclusion}\label{conclusion}
This study introduced a systematic framework for venous constitutive model discovery. Modified constitutive artificial neural networks, trained on biaxial mechanical testing data, were used to discover the terms needed to reproduce venous mechanical behavior; sensitivity analysis, correlation-based pruning, response-level family collapse, and subject-specific refitting were then used to reduce this basis to a compact and interpretable constitutive representation. Although the individual basis functions selected by the subject-specific CANNs varied, a common form SEF was consistently supported across saphenous, jugular, and inferior vena cava specimens.

The resulting eight-term SEF combines an effective isotropic response, direction-specific fiber contributions, recruitment-modified diagonal and circumferential responses, interaction between the diagonal fiber families, and direct circumferential--axial coupling. The reduced formulation retained the in-sample fitting performance of a 16-term parent basis with minimal loss of accuracy. Ablation analyses provided the strongest support for fiber recruitment, and diagonal fiber families, whereas the isotropic and direct strain-coupling terms contributed more selectively across specimens. In the common comparison cohort, the proposed SEF also achieved higher overall fitting accuracy and more favorable AIC and BIC values than the quadratic--Fung, quadratic four-fiber-family, and neo-Hookean four-fiber-family models considered in this study.

The primary conclusion is therefore architectural rather than uniquely functional: venous biaxial behavior was described most effectively by a compact model that combines directional recruitment, fiber-family interaction, and biaxial coupling, while the precise nonlinear function assigned to each family remained partly non-identifiable over the available deformation range. The present results establish descriptive adequacy within the fitted dataset, but external validation using held-out loading protocols, additional venous beds, and independent cohorts will be required before the model can be considered predictive or broadly generalizable. With such validation, the proposed form of the SEF could provide a practical constitutive law for simulations of vein grafts, arteriovenous fistulas, and tissue-engineered vascular grafts.

\backmatter

\bmhead{Acknowledgements}
This work was supported by the National Institutes of Health (NIH) under project numbers R01 HL159954, 700284-0324-00 to A.L.M, and NHLBI T32HL166155 to M.S.B. We thank David Li for helpful discussions that motivated the fiber recruitment formulation used in this work, building on his previous research \cite{li_multi-scale_2023}. We also thank Abhay Ramachandra for sharing his published jugular vein biaxial data \cite{jugular_vein_dataset_Ramachandra2022-jy}





\bibliography{sn-bibliography}

@ARTICLE{Ferruzzi2013-mg_exp_method,
  title     = "Biomechanical phenotyping of central arteries in health and
               disease: advantages of and methods for murine models",
  author    = "Ferruzzi, J and Bersi, M R and Humphrey, J D",
  journal   = "Ann. Biomed. Eng.",
  publisher = "Springer Science and Business Media LLC",
  volume    =  41,
  number    =  7,
  pages     = "1311--1330",
  month     =  jul,
  year      =  2013,
  language  = "en"
}

@ARTICLE{Linka2023-ka_CANN_constraints,
  title     = "A new family of Constitutive Artificial Neural Networks towards
               automated model discovery",
  author    = "Linka, Kevin and Kuhl, Ellen",
  journal   = "Comput. Methods Appl. Mech. Eng.",
  publisher = "Elsevier BV",
  volume    =  403,
  number    =  115731,
  pages     = "115731",
  month     =  jan,
  year      =  2023,
  copyright = "http://creativecommons.org/licenses/by/4.0/",
  language  = "en"
}

@ARTICLE{Akaike1974-co_AIC,
  title     = "A new look at the statistical model identification",
  author    = "Akaike, H",
  journal   = "IEEE Trans. Automat. Contr.",
  publisher = "Institute of Electrical and Electronics Engineers (IEEE)",
  volume    =  19,
  number    =  6,
  pages     = "716--723",
  month     =  dec,
  year      =  1974,
  copyright = "https://ieeexplore.ieee.org/Xplorehelp/downloads/license-information/IEEE.html",
  language  = "en"
}

@article{BIC_c4048c8f-6ca9-3965-96a3-653ab8996955,
 ISSN = {00905364, 21688966},
 URL = {http://www.jstor.org/stable/2958889},
 author = {Gideon Schwarz},
 journal = {The Annals of Statistics},
 number = {2},
 pages = {461--464},
 publisher = {Institute of Mathematical Statistics},
 title = {Estimating the Dimension of a Model},
 urldate = {2026-06-29},
 volume = {6},
 year = {1978}
}

@misc{peirlinck_universal_2024,
    title = {A universal material model subroutine for soft matter systems},
    url = {http://arxiv.org/abs/2404.13144},
    language = {en},
    urldate = {2024-05-29},
    publisher = {arXiv},
    author = {Peirlinck, Mathias and Hurtado, Juan A. and Rausch, Manuel K. and Tepole, Adrian Buganza and Kuhl, Ellen},
    month = apr,
    year = {2024},
}

@article{linka_constitutive_2021,
    title = {Constitutive artificial neural networks: {A} fast and general approach to predictive data-driven constitutive modeling by deep learning},
    volume = {429},
    issn = {00219991},
    shorttitle = {Constitutive artificial neural networks},
    url = {https://linkinghub.elsevier.com/retrieve/pii/S0021999120307841},
    doi = {10.1016/j.jcp.2020.110010},
    language = {en},
    urldate = {2025-09-25},
    journal = {Journal of Computational Physics},
    author = {Linka, Kevin and Hillgärtner, Markus and Abdolazizi, Kian P. and Aydin, Roland C. and Itskov, Mikhail and Cyron, Christian J.},
    month = mar,
    year = {2021},
    pages = {110010},
}

@article{li_multi-scale_2023,
    title = {A multi-scale computational model for the passive mechanical behavior of right ventricular myocardium},
    volume = {142},
    issn = {17516161},
    url = {https://linkinghub.elsevier.com/retrieve/pii/S1751616123001418},
    doi = {10.1016/j.jmbbm.2023.105788},
    language = {en},
    urldate = {2025-10-08},
    journal = {Journal of the Mechanical Behavior of Biomedical Materials},
    author = {Li, David S. and Mendiola, Emilio A. and Avazmohammadi, Reza and Sachse, Frank B. and Sacks, Michael S.},
    month = jun,
    year = {2023},
    pages = {105788},
}

@misc{pierre_principal-stretch-based_2023,
    title = {Principal-stretch-based constitutive neural networks autonomously discover a subclass of {Ogden} models for human brain tissue},
    copyright = {© 2023, Posted by Cold Spring Harbor Laboratory. This pre-print is available under a Creative Commons License (Attribution 4.0 International), CC BY 4.0, as described at http://creativecommons.org/licenses/by/4.0/},
    url = {https://www.biorxiv.org/content/10.1101/2023.01.14.524079v1},
    doi = {10.1101/2023.01.14.524079},
    language = {en},
    urldate = {2026-04-06},
    publisher = {bioRxiv},
    author = {Pierre, Sarah R. St and Linka, Kevin and Kuhl, Ellen},
    month = jan,
    year = {2023},
    note = {Pages: 2023.01.14.524079
Section: New Results},
}

@article{desai_inhibition_2010,
    address = {London, England},
    title = {Inhibition of neointimal formation and hyperplasia in vein grafts by external stent/sheath},
    volume = {15},
    issn = {1477-0377},
    doi = {10.1177/1358863X10366479},
    language = {eng},
    number = {4},
    journal = {Vascular Medicine},
    author = {Desai, Mital and Mirzay-Razzaz, Jalaledin and von Delft, Dirk and Sarkar, Sandip and Hamilton, George and Seifalian, Alexander M.},
    month = aug,
    year = {2010},
    pages = {287--297},
}

@article{mattson_structural_2017,
    title = {Structural and {Functional} {Differences} {Between} {Porcine} {Aorta} and {Vena} {Cava}},
    volume = {139},
    issn = {0148-0731},
    url = {https://pmc.ncbi.nlm.nih.gov/articles/PMC5467030/},
    doi = {10.1115/1.4036261},
    number = {7},
    urldate = {2026-01-26},
    journal = {Journal of Biomechanical Engineering},
    author = {Mattson, Jeffrey M. and Zhang, Yanhang},
    month = jul,
    year = {2017},
    pages = {0710071--0710078},
}

@article{owens_lower_2008,
    title = {Lower extremity vein graft failure: a translational approach},
    volume = {13},
    issn = {1358-863X},
    shorttitle = {Lower extremity vein graft failure},
    url = {https://doi.org/10.1177/1358863X07083432},
    doi = {10.1177/1358863X07083432},
    number = {1},
    urldate = {2026-04-03},
    journal = {Vascular Medicine},
    publisher = {SAGE Publications Ltd STM},
    author = {Owens, Christopher D and Ho, Karen J and Conte, Michael S},
    month = feb,
    year = {2008},
    pages = {63--74},
}

@article{greenwald_improving_2000,
    title = {Improving vascular grafts: the importance of mechanical and haemodynamic properties},
    volume = {190},
    copyright = {Copyright © 2000 John Wiley \& Sons, Ltd.},
    issn = {1096-9896},
    shorttitle = {Improving vascular grafts},
    url = {https://onlinelibrary.wiley.com/doi/abs/10.1002/%28SICI%291096-9896%28200002%29190%3A3%3C292%3A%3AAID-PATH528%3E3.0.CO%3B2-S},
    doi = {10.1002/(SICI)1096-9896(200002)190:3<292::AID-PATH528>3.0.CO;2-S},
    language = {en},
    number = {3},
    urldate = {2026-04-03},
    journal = {The Journal of Pathology},
    author = {Greenwald, S. E. and Berry, C. L.},
    year = {2000},
    note = {\_eprint: https://pathsocjournals.onlinelibrary.wiley.com/doi/pdf/10.1002/\%28SICI\%291096-9896\%28200002\%29190\%3A3\%3C292\%3A\%3AAID-PATH528\%3E3.0.CO\%3B2-S},
    pages = {292--299},
}

@article{prim_evaluation_2021,
    title = {Evaluation of the {Stress}–{Growth} {Hypothesis} in {Saphenous} {Vein} {Perfusion} {Culture}},
    volume = {49},
    issn = {1573-9686},
    url = {https://doi.org/10.1007/s10439-020-02582-1},
    doi = {10.1007/s10439-020-02582-1},
    language = {en},
    number = {1},
    urldate = {2026-04-03},
    journal = {Annals of Biomedical Engineering},
    author = {Prim, David A. and Lane, Brooks A. and Ferruzzi, Jacopo and Shazly, Tarek and Eberth, John F.},
    month = jan,
    year = {2021},
    pages = {487--501},
}

@article{salacinski_mechanical_2001,
    title = {The {Mechanical} {Behavior} of {Vascular} {Grafts}: {A} {Review}},
    volume = {15},
    issn = {0885-3282},
    shorttitle = {The {Mechanical} {Behavior} of {Vascular} {Grafts}},
    url = {https://doi.org/10.1106/NA5T-J57A-JTDD-FD04},
    doi = {10.1106/NA5T-J57A-JTDD-FD04},
    language = {EN},
    number = {3},
    urldate = {2026-04-03},
    journal = {Journal of Biomaterials Applications},
    publisher = {SAGE Publications Ltd STM},
    author = {Salacinski, Henryk J. and Goldner, Sean and Giudiceandrea, Alberto and Hamilton, George and Seifalian, Alexander M. and Edwards, Alan and Carson, Robert J.},
    month = jan,
    year = {2001},
    pages = {241--278},
}

@article{abbott_effect_1987,
    title = {Effect of compliance mismatch on vascular graft patency},
    volume = {5},
    issn = {0741-5214, 1097-6809},
    url = {https://www.jvascsurg.org/article/0741-5214(87)90148-0/fulltext},
    doi = {10.1016/0741-5214(87)90148-0},
    language = {English},
    number = {2},
    urldate = {2026-04-03},
    journal = {Journal of Vascular Surgery},
    publisher = {Elsevier},
    author = {Abbott, William M. and Megerman, Joseph and Hasson, Jonathan E. and L'Italien, Gilbert and Warnock, David F.},
    month = feb,
    year = {1987},
    pages = {376--382},
}

@article{yao_vivo_2023,
    title = {In vivo evaluation of compliance mismatch on intimal hyperplasia formation in small diameter vascular grafts},
    volume = {11},
    issn = {2047-4849},
    url = {https://pubs.rsc.org/en/content/articlelanding/2023/bm/d3bm00167a},
    doi = {10.1039/D3BM00167A},
    language = {en},
    number = {9},
    urldate = {2026-04-03},
    journal = {Biomaterials Science},
    publisher = {The Royal Society of Chemistry},
    author = {Yao, Yuan and Pohan, Grace and Cutiongco, Marie F. A. and Jeong, YeJin and Kunihiro, Joshua and Zaw, Aung Moe and David, Dency and Shangguan, Hanyue and Yu, Alfred C. H. and Yim, Evelyn K. F.},
    month = may,
    year = {2023},
    pages = {3297--3307},
}

@article{ballyk_compliance_1997,
    title = {Compliance mismatch may promote graft–artery intimal hyperplasia by altering suture-line stresses},
    volume = {31},
    issn = {0021-9290},
    url = {https://www.sciencedirect.com/science/article/pii/S0197397597001115},
    doi = {10.1016/S0197-3975(97)00111-5},
    number = {3},
    urldate = {2026-04-03},
    journal = {Journal of Biomechanics},
    author = {Ballyk, Peter D and Walsh, Colin and Butany, Jagadish and Ojha, Matadial},
    month = dec,
    year = {1997},
    pages = {229--237},
}

@article{stewart_effects_1992,
    title = {Effects of a vascular graft/natural artery compliance mismatch on pulsatile flow},
    volume = {25},
    issn = {0021-9290},
    url = {https://www.sciencedirect.com/science/article/pii/002192909290027X},
    doi = {10.1016/0021-9290(92)90027-X},
    number = {3},
    urldate = {2026-04-03},
    journal = {Journal of Biomechanics},
    author = {Stewart, Sandy F. C. and Lyman, Donald J.},
    month = mar,
    year = {1992},
    pages = {297--310},
}

@article{smith_impact_2016,
    title = {The {Impact} of {Vein} {Mechanical} {Compliance} on {Arteriovenous} {Fistula} {Outcomes}},
    volume = {32},
    issn = {0890-5096, 1615-5947},
    url = {https://www.annalsofvascularsurgery.com/article/S0890-5096(16)00027-3/abstract},
    doi = {10.1016/j.avsg.2015.11.002},
    language = {English},
    urldate = {2026-04-03},
    journal = {Annals of Vascular Surgery},
    publisher = {Elsevier},
    author = {Smith, George E. and Barnes, Rachel and Fagan, Michael and Chetter, Ian C.},
    month = apr,
    year = {2016},
    pages = {9--14},
}

@article{kao_constitutive_2010,
    title = {Constitutive {Modeling} of {Anisotropic} {Finite}-{Deformation} {Hyperelastic} {Behaviors} of {Soft} {Materials} {Reinforced} by {Tortuous} {Fibers}.},
    url = {https://www.semanticscholar.org/paper/Constitutive-Modeling-of-Anisotropic-Hyperelastic-Kao-Lammers/6ece06b9328baa9635d101774b85bd81dcdf2c96},
    urldate = {2026-04-03},
    journal = {The international journal of structural changes in solids : mechanics and applications},
    author = {Kao, P. and Lammers, Steven R. and Hunter, K. and Stenmark, K. and Shandas, R. and Qi, H.},
    month = apr,
    year = {2010},
}

@article{anssari-benam_specialized_2020,
    title = {Specialized {Strain} {Energy} {Functions} for {Modeling} the {Contribution} of the {Collagen} {Network} ({Waniso}) to the {Deformation} of {Soft} {Tissues}},
    volume = {87},
    issn = {0021-8936},
    url = {https://doi.org/10.1115/1.4046894},
    doi = {10.1115/1.4046894},
    number = {071006},
    urldate = {2026-04-03},
    journal = {Journal of Applied Mechanics},
    author = {Anssari-Benam, Afshin and Pani, Martino and Bucchi, Andrea},
    month = may,
    year = {2020},
}

@article{saez_microstructural_2016,
    title = {Microstructural quantification of collagen fiber orientations and its integration in constitutive modeling of the porcine carotid artery},
    volume = {33},
    issn = {1742-7061},
    url = {https://www.sciencedirect.com/science/article/pii/S1742706116300307},
    doi = {10.1016/j.actbio.2016.01.030},
    urldate = {2026-04-03},
    journal = {Acta Biomaterialia},
    author = {Sáez, P. and García, A. and Peña, E. and Gasser, T. C. and Martínez, M. A.},
    month = mar,
    year = {2016},
    pages = {183--193},
}

@article{sokolis_experimental_2013,
    title = {Experimental investigation and constitutive modeling of the {3D} histomechanical properties of vein tissue},
    volume = {12},
    issn = {1617-7940},
    url = {https://doi.org/10.1007/s10237-012-0410-y},
    doi = {10.1007/s10237-012-0410-y},
    language = {en},
    number = {3},
    urldate = {2026-04-22},
    journal = {Biomechanics and Modeling in Mechanobiology},
    author = {Sokolis, Dimitrios P.},
    month = jun,
    year = {2013},
    pages = {431--451},
}

@book{humphrey_cardiovascular_2002,
    address = {New York, NY},
    title = {Cardiovascular {Solid} {Mechanics}: {Cells}, {Tissues}, and {Organs}},
    copyright = {https://www.springernature.com/gp/researchers/text-and-data-mining},
    isbn = {978-1-4419-2897-9 978-0-387-21576-1},
    shorttitle = {Cardiovascular {Solid} {Mechanics}},
    url = {https://link.springer.com/10.1007/978-0-387-21576-1},
    doi = {10.1007/978-0-387-21576-1},
    language = {en},
    urldate = {2026-03-31},
    publisher = {Springer},
    author = {Humphrey, Jay D.},
    year = {2002},
}

@phdthesis{saitta-rezakhaniha_biomechanics_2010,
    title = {Biomechanics of {Vascular} {Wall} : the {Role} of {Structural} {Organization} of {Elastin} and {Collagen}},
    shorttitle = {Biomechanics of {Vascular} {Wall}},
    url = {https://infoscience.epfl.ch/handle/20.500.14299/49046},
    doi = {10.5075/epfl-thesis-4712},
    language = {en},
    urldate = {2026-04-03},
    school = {EPFL},
    author = {Saitta-Rezakhaniha, Rana},
    year = {2010},
    doi = {10.5075/epfl-thesis-4712},
}

@ARTICLE{Famaey2015-vk_review_of_bridging_veins,
  title     = "Structural and mechanical characterisation of bridging veins: A
               review",
  author    = "Famaey, Nele and Ying Cui, Zhao and Umuhire Musigazi, Grace and
               Ivens, Jan and Depreitere, Bart and Verbeken, Erik and Vander
               Sloten, Jos",
  journal   = "J. Mech. Behav. Biomed. Mater.",
  publisher = "Elsevier BV",
  volume    =  41,
  pages     = "222--240",
  month     =  jan,
  year      =  2015,
  language  = "en"
}

@ARTICLE{Lee2013-ao,
  title     = "Biaxial mechanical properties of the inferior vena cava in
               {C57BL/6} and {CB-17} {SCID/bg} mice",
  author    = "Lee, Y U and Naito, Y and Kurobe, H and Breuer, C K and
               Humphrey, J D",
  journal   = "J. Biomech.",
  publisher = "Elsevier BV",
  volume    =  46,
  number    =  13,
  pages     = "2277--2282",
  month     =  sep,
  year      =  2013,
  language  = "en"
}

@ARTICLE{Desch2007-is,
  title     = "A model for passive elastic properties of rat vena cava",
  author    = "Desch, Georg Wolfgang and Weizs{\"a}cker, Hans Werner",
  journal   = "J. Biomech.",
  publisher = "Elsevier BV",
  volume    =  40,
  number    =  14,
  pages     = "3130--3145",
  month     =  may,
  year      =  2007,
  language  = "en"
}

@ARTICLE{Rezakhaniha2008-nh,
  title     = "A structural model of the venous wall considering elastin
               anisotropy",
  author    = "Rezakhaniha, Rana and Stergiopulos, Nikos",
  journal   = "J. Biomech. Eng.",
  publisher = "ASME International",
  volume    =  130,
  number    =  3,
  pages     = "031017",
  month     =  jun,
  year      =  2008,
  language  = "en"
}

@article{sobol_global_2001,
    series = {The {Second} {IMACS} {Seminar} on {Monte} {Carlo} {Methods}},
    title = {Global sensitivity indices for nonlinear mathematical models and their {Monte} {Carlo} estimates},
    volume = {55},
    issn = {0378-4754},
    url = {https://www.sciencedirect.com/science/article/pii/S0378475400002706},
    doi = {10.1016/S0378-4754(00)00270-6},
    number = {1},
    urldate = {2026-08-03},
    journal = {Mathematics and Computers in Simulation},
    author = {Sobol′, I. M},
    month = feb,
    year = {2001},
    pages = {271--280},
}

@article{saltelli_variance_2010,
    title = {Variance based sensitivity analysis of model output. {Design} and estimator for the total sensitivity index},
    volume = {181},
    issn = {0010-4655},
    url = {https://www.sciencedirect.com/science/article/pii/S0010465509003087},
    doi = {10.1016/j.cpc.2009.09.018},
    number = {2},
    urldate = {2026-08-03},
    journal = {Computer Physics Communications},
    author = {Saltelli, Andrea and Annoni, Paola and Azzini, Ivano and Campolongo, Francesca and Ratto, Marco and Tarantola, Stefano},
    month = feb,
    year = {2010},
    pages = {259--270},
}

@ARTICLE{hgo_model_Gasser2006-vd,
  title     = "Hyperelastic modelling of arterial layers with distributed
               collagen fibre orientations",
  author    = "Gasser, T Christian and Ogden, Ray W and Holzapfel, Gerhard A",
  journal   = "J. R. Soc. Interface",
  publisher = "The Royal Society",
  volume    =  3,
  number    =  6,
  pages     = "15--35",
  month     =  feb,
  year      =  2006,
  language  = "en"
}

@ARTICLE{ellipticity_Wilber2002-wt,
  title     = "The convexity properties of a class of constitutive models for
               biological soft issues",
  author    = "Wilber, J Patrick and Walton, Jay R",
  journal   = "Math. Mech. Solids",
  publisher = "SAGE Publications",
  volume    =  7,
  number    =  3,
  pages     = "217--235",
  month     =  jun,
  year      =  2002,
  language  = "en"
}

@ARTICLE{polyconvexity_Klein2026-ja,
  title         = "On limitations of polyconvexity",
  author        = "Klein, Dominik K and Ortigosa, Rogelio and Roth, Heinrich T
                   and Kalina, Karl A and Mart{\'\i}nez-Frutos, Jes{\'u}s and
                   K{\"a}stner, Markus and Weeger, Oliver",
  month         =  may,
  year          =  2026,
  copyright     = "http://creativecommons.org/licenses/by/4.0/",
  archivePrefix = "arXiv",
  primaryClass  = "cs.CE",
  eprint        = "2605.31392"
}

@ARTICLE{jugular_vein_dataset_Ramachandra2022-jy,
  title     = "Biodegradable external wrapping promotes favorable adaptation in
               an ovine vein graft model",
  author    = "Ramachandra, Abhay B and Wang, Hanjay and Wnorowski, Alexa and
               Schwarz, Erica L and Pickering, Joshua and Heiler, Joseph C and
               Lucian, Haley J and Hironaka, Camille E and Tran, Nicholas A and
               Liu, Yu and Khan, Muhammad Owais and Obafemi, Oluwatomisin and
               Tada, Yuko and Kahn, Andrew M and Sayed, Nazish and Wu, Joseph C
               and Humphrey, Jay D and Boyd, Jack H and Marsden, Alison L",
  journal   = "Acta Biomater.",
  publisher = "Elsevier BV",
  volume    =  151,
  pages     = "414--425",
  month     =  oct,
  year      =  2022,
  language  = "en"
}

@ARTICLE{vessels_comparison_Ramachandra2024-je,
  title     = "A systematic comparison of normal structure and function of the
               greater thoracic vessels",
  author    = "Ramachandra, Abhay B and Cavinato, Cristina and Humphrey, Jay D",
  journal   = "Ann. Biomed. Eng.",
  publisher = "Springer Science and Business Media LLC",
  volume    =  52,
  number    =  4,
  pages     = "958--966",
  month     =  apr,
  year      =  2024,
  copyright = "https://www.springernature.com/gp/researchers/text-and-data-mining",
  language  = "en"
}

@ARTICLE{CANN_Linka2022-ak,
  title     = "Predicting and understanding arterial elasticity from key
               microstructural features by bidirectional deep learning",
  author    = "Linka, Kevin and Cavinato, Cristina and Humphrey, Jay D and
               Cyron, Christian J",
  journal   = "Acta Biomater.",
  publisher = "Elsevier BV",
  volume    =  147,
  pages     = "63--72",
  month     =  jul,
  year      =  2022,
  copyright = "http://creativecommons.org/licenses/by-nc-nd/4.0/",
  language  = "en"
}

@article{hart_robustness_2019,
    title = {Robustness of the {Sobol}' {Indices} to {Marginal} {Distribution} {Uncertainty}},
    volume = {7},
    issn = {2166-2525},
    url = {https://epubs.siam.org/doi/10.1137/18M123387X},
    doi = {10.1137/18M123387X},
    language = {en},
    number = {4},
    urldate = {2026-08-18},
    journal = {SIAM/ASA Journal on Uncertainty Quantification},
    author = {Hart, Joseph L. and Gremaud, Pierre A.},
    month = jan,
    year = {2019},
    pages = {1224--1244},
}

@ARTICLE{Humphrey_torque_Emuna2026-zn,
  title    = "Optimized biomechanical design of a tissue engineered pulsatile
              Fontan conduit",
  author   = "Emuna, Nir and Hu, Zinan and Marsden, Alison L and Humphrey, Jay
              D",
  journal  = "NPJ Regen. Med.",
  volume   =  11,
  number   =  1,
  pages    = "6",
  month    =  jan,
  year     =  2026,
  language = "en"
}

@ARTICLE{Baek2007-dw,
  title     = "Theory of small on large: Potential utility in computations of
               fluid--solid interactions in arteries",
  author    = "Baek, S and Gleason, R L and Rajagopal, K R and Humphrey, J D",
  journal   = "Comput. Methods Appl. Mech. Eng.",
  publisher = "Elsevier BV",
  volume    =  196,
  number    = "31-32",
  pages     = "3070--3078",
  month     =  jun,
  year      =  2007,
  language  = "en"
}

\clearpage

\begin{appendices}
\renewcommand{\thesubsection}{S.\arabic{subsection}}
\setcounter{subsection}{0}
\renewcommand{\thetable}{S\arabic{table}}
\renewcommand{\thefigure}{S\arabic{figure}}
\setcounter{table}{0}
\setcounter{figure}{0}
\renewcommand{\thefigure}{S\arabic{figure}}
\renewcommand{\figurename}{Supplementary Figure}

\section*{Supplementary Information}
\subsection{Additional Methods Details}
\subsubsection{Loss function and training} \label{loss}
Each of the nine tissue samples was treated as an independent dataset during the constitutive modeling discovery, and a separate CANN was trained for each specimen without pooling mechanical data across specimens or vein type. For each specimen, the circumferential and axial stress--stretch observations from the three pressure--diameter protocols and four force--length protocols listed in Table~\ref{tab:training_points_by_vein} were combined and fitted simultaneously. The model predicted the strain energy density function, which was differentiated with respect to the deformation to obtain the Cauchy stress components, $\hat{\sigma}_{\theta}$ and $\hat{\sigma}_{z}$. 

Because the pressure--diameter and force--length protocols contained different numbers of data points and exhibited different stress magnitudes, the stress residuals were normalized independently in the circumferential and axial directions. The loss was defined as
\begin{equation}
\mathcal{L}_{\mathrm{data}} =\frac{1}{N_\theta}\sum_{i=1}^{N_\theta}\omega_{\theta,i}\rho\left(\frac{\hat{\sigma}_{\theta,i}-\sigma_{\theta,i}}{s_\theta}\right)+\frac{1}{N_z}\sum_{i=1}^{N_z}\omega_{z,i}\rho\left(\frac{\hat{\sigma}_{z,i}-\sigma_{z,i}}{s_z}\right),
\label{eq:data_loss}
\end{equation}
where $N_\theta$ and $N_z$ are the numbers of circumferential and axial stress data points, respectively; $\omega_{\theta,i}$ and $\omega_{z,i}$ are sample-specific weights; $\rho(\cdot)$ is a robust residual penalty; and $s_\theta$ and $s_z$ are stress normalization factors defined as
\begin{equation}
s_\theta = \max_i \left| \sigma_{\theta,i} \right|,\qquad s_z
= \max_i\left| \sigma_{z,i}\right|.
\end{equation}
We include two types of penalties: Huber loss and mean squared error loss. For the Huber loss used in the robust training configuration, the penalty becomes:
\begin{equation}
\rho_{\mathrm{Huber}}(r)
=
\begin{cases}
\dfrac{1}{2}r^2,
& |r| < \delta, \\[6pt]
\delta\left(|r|-\dfrac{1}{2}\delta\right),
& |r| \geq \delta.
\end{cases}
\label{eq:huber_loss}
\end{equation}
where $r$ denotes the normalized stress residual and $\delta$ controls the transition between the quadratic and linear regimes. Thus, small residuals were penalized quadratically, whereas large residuals were penalized approximately linearly. This retained sensitivity to the overall stress--stretch response while limiting the influence of potential experimental outliers. When the standard mean-squared-error configuration was used, $\rho(r)=r^2$.

The sample weights were expressed as
\begin{equation}
\omega_{\theta,i} = w_\theta \omega_{\theta,i}^{\mathrm{curve}} \omega_{\theta,i}^{\mathrm{tail}}, \qquad \omega_{z,i} = w_z \omega_{z,i}^{\mathrm{curve}} \omega_{z,i}^{\mathrm{tail}},
\end{equation}
where $w_\theta$ and $w_z$ controlled the relative contributions of the circumferential and axial responses. The curve-level weights, $\omega^{\mathrm{curve}}$, were defined such that each loading protocol contributed equally to the loss regardless of the number of recorded points. The tail weights, $\omega^{\mathrm{tail}}$, increased the contribution of observations near the transition from the compliant low-stiffness response to the stiffened high-stiffness response. 

The total training objective included the standard $L1$ and $L2$ regularization penalties added to the loss defined in eq. \ref{eq:data_loss}. Model parameters were optimized using the Adam algorithm. The learning rate was reduced when the monitored loss reached a plateau, and early stopping with restoration of the best-performing weights was used to limit overfitting.

After training, candidate terms with negligible learned weights or minimal contribution to the predicted stress response were removed from the retained library. The surviving terms constituted the discovered SEF for each specimen, which was carried forward for sensitivity analysis, correlation-based redundancy filtering, and reduced-basis refitting.

\subsubsection{Sensitivity analysis and Sobol ranking} \label{sup_sobol-selection}
The discovered subject-specific SEFs, were pooled into a candidate library comprising the union of all terms active across the nine trained samples. While this pooled library offered sufficient flexibility to represent nonlinear, anisotropic, and coupled deformation responses, it also contained terms with potentially overlapping mechanical effects. Therefore, we performed a variance-based Sobol sensitivity analysis to rank their influence \cite{sobol_global_2001, saltelli_variance_2010}.

Each candidate term, represented by basis function $\psi_j$, was evaluated over two biaxial stretch domains: physiological and extended grids. The physiological grid consists of a $20\times20$ uniform grid spanning $0.80 \leq \lambda_\theta \leq 1.60$ and $1.00 \leq \lambda_z \leq 1.40$, corresponding to the deformation range most relevant to the experimental biaxial data. The extended grid also consist of a $20\times20$ uniform grid, but covering a broader domain $0.30 \leq \lambda_\theta \leq 2.10$ and $1.00 \leq \lambda_z \leq 1.60$. The physiological characterized term influence within the experimentally relevant range, whereas the extended grid was used to examine whether term influence remained consistent over a broader deformation domain.

For each domain $g\in{\mathrm{phys},\mathrm{ext}}$, the candidate basis $N_b$ was evaluated at $M=400$ biaxial stretch states to produce the basis matrix $\boldsymbol{\psi}$
\begin{equation}
    \boldsymbol{\psi^g} = \begin{bmatrix}
\psi_1\!\left(\lambda_{\theta,1}^{(g)},\lambda_{z,1}^{(g)}\right)
& \cdots &
\psi_{N_b}\!\left(\lambda_{\theta,1}^{(g)},\lambda_{z,1}^{(g)}\right)
\\
\vdots & \ddots & \vdots
\\
\psi_1\!\left(\lambda_{\theta,M}^{(g)},\lambda_{z,M}^{(g)}\right)
& \cdots &
\psi_{N_b}\!\left(\lambda_{\theta,M}^{(g)},\lambda_{z,M}^{(g)}\right)
\end{bmatrix}
\in \mathbb{R}^{M\times N_b},
\label{eq:basis_matrix}
\end{equation}
Each row of \(\boldsymbol{\psi}^{(g)}\) corresponds to one biaxial stretch state, whereas the (j)th column, ~$\boldsymbol{\psi}_j^{(g)}$, contains the energy response of one candidate basis function across all \(M\) states.
For a sampled coefficient vector $\mathbf{\alpha}=[\alpha_1,\ldots,\alpha_{N_b}]$, the strain-energy density values across domain $g$  were calculated as
\begin{equation}
W^g(\lambda_\theta,\lambda_z;\mathbf{\alpha})=\sum_{j=1}^{N_b}\alpha_j\psi_j^g(\lambda_\theta,\lambda_z),
\label{eq:sobol_energy_field}
\end{equation}
where \(\alpha_j\) are the sampled weight coefficients to the candidate function \(\psi_j\), and $N_b$ is the total number of candidate basis functions over the stretch grid for every $\mathbf{w}$. 

Since Sobol sensitivity indices depend both on the numerical magnitude of a basis function and on the range over which its coefficient is sampled,  the primary analysis used a the \emph{normalized-cohort} regime ~\cite{hart_robustness_2019}. In this regime, each basis-function column was divided by its Euclidean norm to reduce differences in numerical scale among candidate terms. And its coefficient were sampled from $[-s_j, s_j]$, where $s_j$ is the cross-animal standard deviation of the corresponding trained effective coefficient. These zero-centered ranges were used only for sensitivity screening, not as physically admissible coefficient bounds, so the resulting indices quantify sensitivity to observed cohort variability.

A scalar model output is required for the Sobol analysis. The energy vector was therefore reduced to its root-mean-square magnitude,
\begin{equation}
Y(\mathbf{\alpha})= \left[\frac{1}{M}\sum_{m=1}^{M} W\left(\lambda_{\theta,m},\lambda_{z,m};\mathbf{w}\right)^2\right]^{1/2},
\label{eq:sobol_rms_energy}
\end{equation}
The scalar response $Y(\mathbf{\alpha})$ thereby represents the overall magnitude of the strain-energy response associated with a given coefficient vector, rather than the energy at a single deformation state.

We decomposed the variance of $Y$ across the sampled coefficients vector $\mathbf{\alpha}$ using first-order $S_j$, and total-order Sobol $S_{T,j}$ indices \cite{sobol_global_2001, saltelli_variance_2010}. The first-order index quantifies the fraction of the variance in the root mean squared energy response associated to variations in $\alpha_j$ alone, whereas the total-order index additionally captures interaction effects involving $\alpha_j$. Candidate terms were ranked using $S_{T,j}$, as the more complete measure of each basis function's influence.

Under the normalized-cohort regime, we calculated $S_{T,j}$ independently on the physiological and extended deformation grids. Within each grid, terms were ranked in descending order of $S_{T,j}$ such that rank 1 corresponded to the largest total-order sensitivity index. The primary Sobol rank for term $j$ was then defined as the mean of its ranks across the two grids,
\begin{equation}
\overline{r}_{T,j} = 
\frac{1}{2}
\left(
r_{T,j}^{\mathrm{phys}}
+
r_{T,j}^{\mathrm{ext}}
\right),
\label{eq:primary_sobol_rank}
\end{equation}
where $r_{T,j}^{\mathrm{phys}}$ and $r_{T,j}^{\mathrm{ext}}$ denote the total-order Sobol ranks obtained on the physiological and extended grids, respectively. Smaller values of $\overline{r}_{T,j}$ indicate a larger index $S_{T,j}$ and thus means that the term was consistently influential across both deformation domains. 

\subsubsection{Correlation-based pruning}\label{sup_methods:cor-pruning}
Sobol analysis identifies terms that influence the overall strain-energy response but does not determine whether two candidate basis produce distinct deformation dependence. We therefore performed a correlation-based pruning to remove mechanically redundant terms.

For each grid $g \in \mathrm{phys}, \mathrm{ext}$, the columns of the matrix $\boldsymbol{\psi^g}$contain the energy response of each candidate function across all \(M\) deformation states. We calculated the absolute Pearson correlation between each pair of evaluated basis function columns and define a cross-grid redundancy metric
\begin{equation}
\rho_{ij}^{\rm{avg}}=\frac{1}{2} \left( \left|\rho_{ij}^{\mathrm{phys}}\right| + 
\left|\rho_{ij}^{\mathrm{ext}}\right|\right),
\label{eq}
\end{equation}
where $\rho_{ij}^{\mathrm{phys}}$ and $\rho_{ij}^{\mathrm{ext}}$ denote the Pearson correlations between the columns for $\psi_i$ and $\psi_j$ on the physiological and extended grids, respectively. Thus $\rho_{ij}^{\rm{avg}}$ identify pairs that exhibited similar deformation dependence, on average, across both domains.

Candidates were considered sequentially in ascending order of primary Sobol rank $\overline{r}_{T,j}$, such that the most influential term was considered first. Each candidate $\psi_j$ was retained only when $ \rho_{ij}^{\rm{avg}} <\rho_{\mathrm{crit}}$ for every previously retained term $i$, where $\rho_{\mathrm{crit}}$ is the prescribed correlation threshold. This greedy, rank-preserving procedure preferentially retains influential terms while limiting redundancy in the design matrix and reducing the likelihood that nearly indistinguishable basis functions share coefficient weight.

\subsubsection{Response-based within-family collapse} \label{sup_methods:family-collapse}
Pairwise correlation-based pruning removed candidate functions that generated nearly identical mechanical responses on both physiological and extended domains. However, correlation-based pruning targeted pairwise numerical redundancy and therefore did not necessarily identify the most compact representation of the mechanical response within each constitutive family. In some cases, several lower-order functions of the same invariant or strain measure were retained because no individual pair exceeded the prescribed correlation threshold, even though their combined response could be represented by a single higher-order function. Response-based within-family reduction was therefore introduced to identify a more compact single-function representation of each retained constitutive family.

Surviving terms were grouped into invariant families, including $I_1$, $I_2$, $I_{4d1}$, $I_{4d2}$, $I_{4z}$, $I_{4\theta}$, and $I_{8d1d2}$, and into pure circumferential, pure axial, and mixed strain families. For each family $\mathcal{F}$, a scalar family variable $x_{\mathcal{F}}$ was defined as
\begin{equation}
x_{\mathcal{F}} =
\begin{cases} I_1 - 3, & \mathcal{F}=I_1, \\[2pt]
I_2 - 3, & \mathcal{F}=I_2, \\[2pt]
I_4 - 1, & \mathcal{F}\in
\left\{ I_{4d2}, I_{4d1}, I_{4z}, I_{4\theta} \right\}, \\[2pt]
I_{8fs},& \mathcal{F}=I_{8d1d2}
\end{cases}
\label{eq:family_variable}
\end{equation}
using the same shifted argument employed in the candidate library. The aggregate response of family $\mathcal{F}$ was reconstructed from the terms surviving decollinearization as
\begin{equation}
\mathcal{F} = 
\sum_{j\in\mathcal{F}}
\overline{c}_j
\psi_j\left(x_{\mathcal{F}}\right),
\label{eq}
\end{equation}
where $\overline{c}_j$ denotes the cohort-mean effective coefficient associated with term $j$, calculated from the subject-specific coefficients obtained from CANN fits. 

For each invariant family, candidate representatives were drawn from the original function library. The tested set included all surviving functions and all same-family library functions with complexity equal to or greater than that of the most complex surviving term. Each candidate,$\widehat{W}_{\mathcal{F}}(x_{\mathcal{F}};\boldsymbol{\theta})$, was fitted to the aggregate family response on the physiological grid using nonlinear least squares.

Multiple initial guesses were used for nonlinear candidates, and the parameters were constrained to remain non-negative. Candidate parameters were fitted on the physiological grid and then evaluated over both the physiological and extended domains. Agreement between each candidate function and the aggregate family response was quantified using the $R^2$ value for the strain energy density function.

\subsubsection{Parameter search methods for reduced SEF} \label{supp_methods_parameter_search_reduced}
The results of family collapse provided a set of possible candidates of SEFs that were evaluated by parameter  estimation to define the best reduced equation. Both linear and nonlinear parameters were estimated separately for each specimen using a nested optimization strategy.  First   $N=4096$ nonlinear parameter vectors were generated by Latin hypercube sampling. For each vector, the stress-basis matrix was evaluated, and the optimal linear coefficients were determined using nonnegative least squares. The nonlinear parameter vector yielding the lowest stress-fitting loss was then used to initialize a bounded local refinement with the limited-memory Broyden--Fletcher--Goldfarb--Shanno algorithm (L-BFGS-B).

The nonlinear search included recruitment parameters for the $d_1$, $d_2$, $z$, and $\theta$ fiber families as described in Section \ref{fiber-recruit}. These parameters consisted of the lower and upper recruitment stretches, $\lambda_{\mathrm{lb}}$ and $\lambda_{\mathrm{ub}}$, the parameters defining the selected recruitment-threshold distribution, and the smoothing parameter $\tau$. The search also included the dispersion parameters $\kappa_f$ and $\kappa_s$; the orientation angles $\phi_{d1}$, $\phi_{d2}$, $\phi_{z}$, and $\phi_{\theta}$; and exponential curvature parameters associated with the $I_{4\mathrm{\theta}}$ and $I_{4z}$ contributions.

Search bounds for each parameter were defined relative to the nonlinear values obtained during subject-specific CANN training, allowing a fixed percentage or absolute variation about the trained value, clipped to a physically admissible interval (Table \ref{tab:nonlinear-search-bounds}). These bounds preserved information learned during CANN training while allowing sufficient variation to refit each compact constitutive equation independently to the subject-specific biaxial stress data.
\begin{table}[t]
\centering
\caption{Bounds used for the nonlinear parameter search of each compact candidate equation. Bounds defined relative to the subject-specific CANN estimates were clipped to the indicated admissible intervals.}
\label{tab:nonlinear-search-bounds}
\small
\setlength{\tabcolsep}{10pt}
\renewcommand{\arraystretch}{1.2}
\begin{tabular}{lcc}
\toprule
\textbf{Parameter} & \textbf{Search bounds} & \textbf{Admissible interval} \\
\midrule
$\lambda_{\mathrm{lb}}$ & $\pm20\%$ & $[0.75,1.80]$ \\
$\lambda_{\mathrm{ub}}$ & $\pm20\%$ & $[\lambda_{\mathrm{lb}}+10^{-3},\,2.40]$ \\
Beta $\alpha,\beta$ & $\pm50\%$ & $[0.25,25]$ \\
Lognormal $\sigma_{\log}$ & $\pm50\%$ & $[0.02,3.0]$ \\
$\tau$ & $\pm50\%$ & $[10^{-4},0.25]$ \\
$\kappa_f,\kappa_s$ & $\pm0.05$ & $[0,1/3]$ \\
Fiber angles $\phi_a$ & $\pm15^\circ$ & angle-specific bounds \\
Curvature parameters & -- & $[10^{-4},5.0]$ \\
\bottomrule
\end{tabular}
\end{table}

\subsection{Results}

\subsubsection{Discovered SEF fits for remaining vein data}\label{supp:remaining_discovered_models}

The discovered SEFs for three representative samples are shown in section \ref{results_cann}. Figures \ref{fig:supp_ivc_fits}, \ref{fig:supp_saphenous_fits} and \ref{fig:supp_jugular_fits} show the discovered fits for the remaining IVC, saphenous and jugular samples, respectively.

\begin{figure}[p]
    \centering
    \includegraphics[width=0.8\linewidth]
    {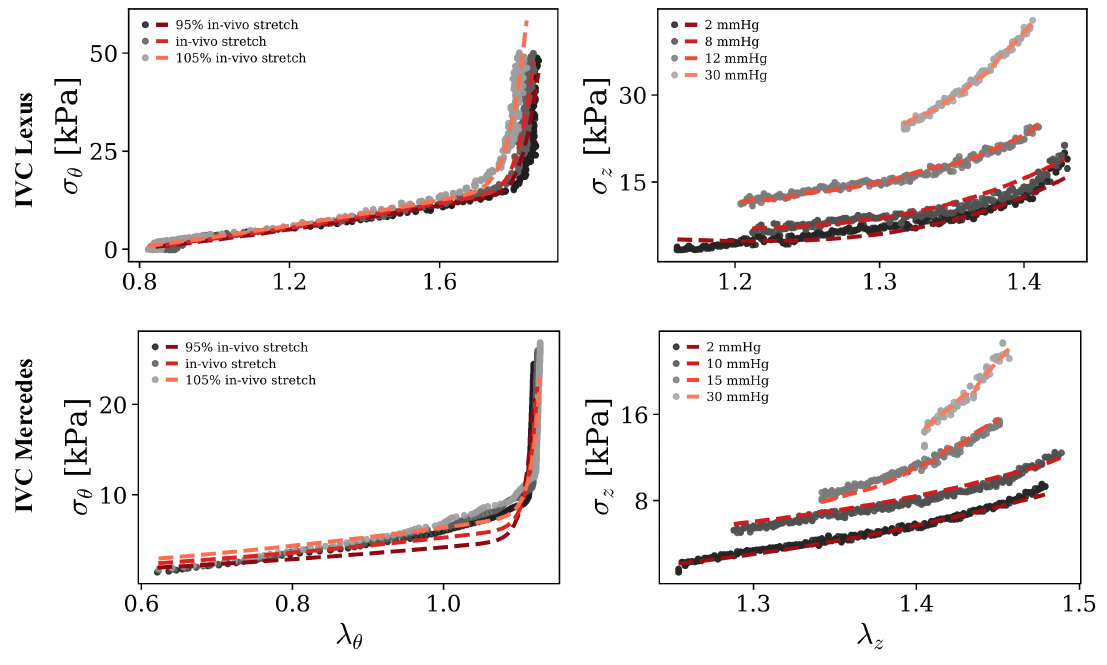}
    \caption{
    Subject-specific CANN fits to biaxial mechanical data from remaining inferior vena cava specimens not shown in \ref{fig:sample_CANN_fit}. Experimental measurements are shown by closed circles, and CANN predictions are shown by dashed curves.
    }
    \label{fig:supp_ivc_fits}
\end{figure}

\begin{figure}[p]
    \centering
    \includegraphics[width=0.8\linewidth]
    {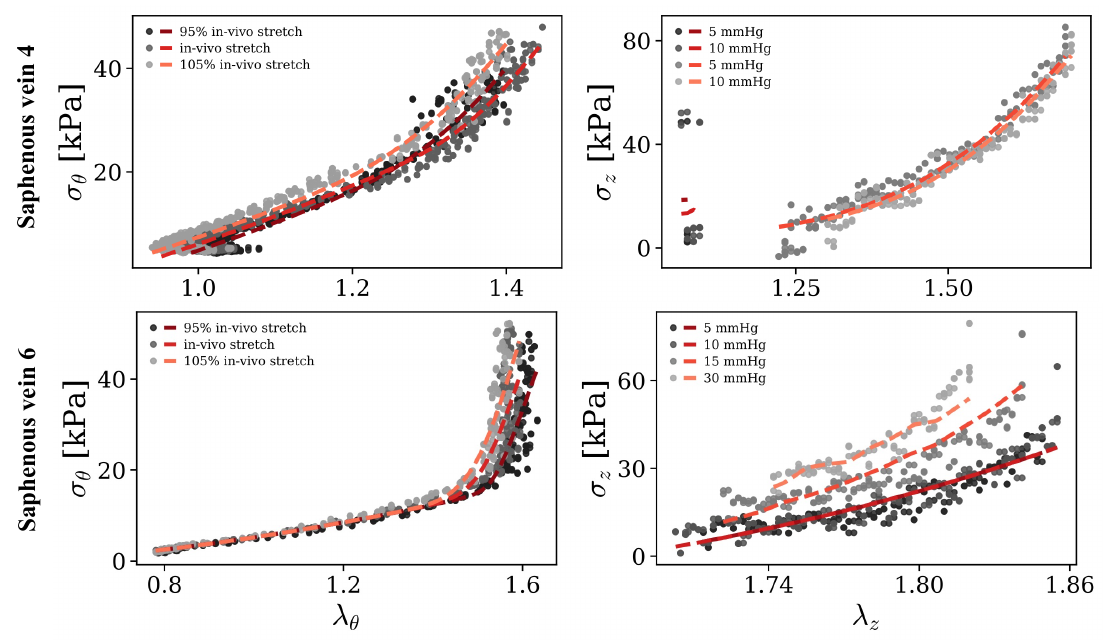}
    \caption{
    Subject-specific CANN fits to biaxial mechanical data from remaining saphenous vein specimens not shown in \ref{fig:sample_CANN_fit}. Experimental measurements are shown by closed circles, and CANN predictions are shown by dashed curves.
    }
    \label{fig:supp_saphenous_fits}
\end{figure}

\begin{figure}[p]
    \centering
    \includegraphics[width=0.8\linewidth]
    {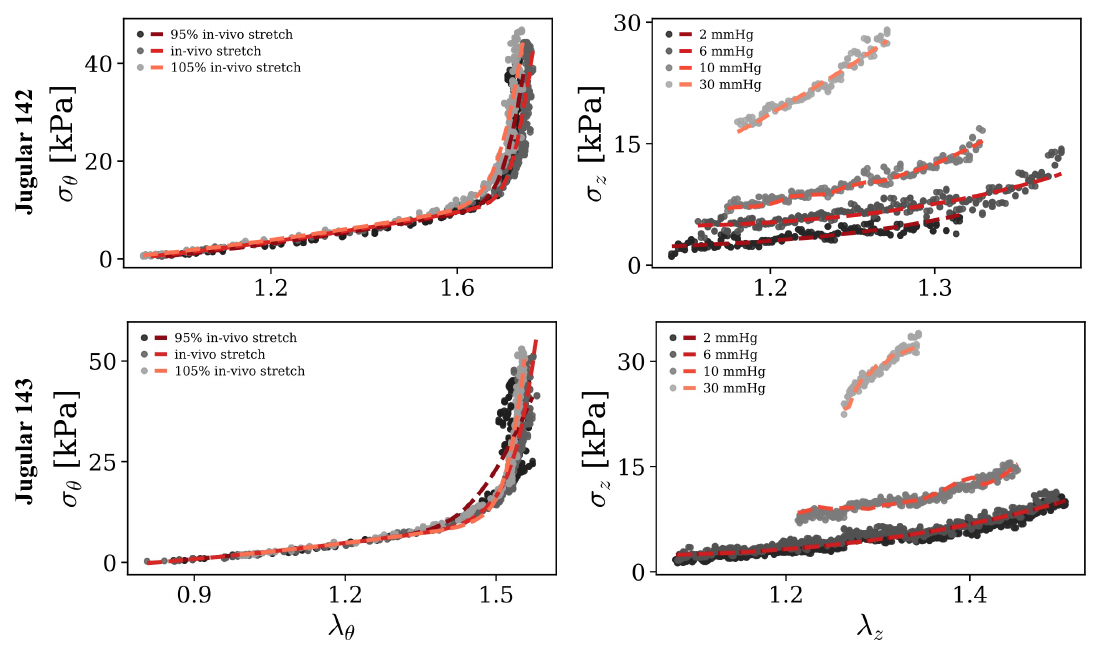}
    \caption{
    Subject-specific CANN fits to biaxial mechanical data from remaining jugular vein specimens not shown in \ref{fig:sample_CANN_fit}. Experimental measurements are shown by closed circles, and CANN predictions are shown by dashed curves.
    }
    \label{fig:supp_jugular_fits}
\end{figure}

\subsubsection{Sensitivity of the discovered candidate basis to the correlation threshold}

Table~\ref{tab:supp-rho-performance} summarizes the effect of the correlation threshold on the number of retained basis functions and the subject-specific fitting performance. Values of $R_w^2$ are reported as the mean $\pm$ standard deviation across smaples. Changes in the cohort mean, $\Delta\overline{R_w^2}$, were calculated relative to the preceding threshold using the unrounded values. The basis obtained at $\rho_{\mathrm{crit}}=0.98$ was retained as an inclusive parent basis for subsequent family collapse rather than as the final constitutive model.
\begin{table}[!htbp]
\centering
\caption{Effect of the correlation threshold on retained basis size and
subject-specific fitting performance.}
\label{tab:supp-rho-performance}

\small
\setlength{\tabcolsep}{8pt}
\renewcommand{\arraystretch}{1.15}

\begin{tabular}{cccc}
\toprule
$\rho_{\mathrm{crit}}$ &
Retained terms &
$\overline{R_w^2}\pm\mathrm{SD}$ &
$\Delta\overline{R_w^2}$ \\
\midrule
0.90 & 8  & $0.878 \pm 0.058$ & -- \\
0.93 & 8  & $0.875 \pm 0.059$ & $-0.003$ \\
0.95 & 8  & $0.875 \pm 0.061$ & $<0.001$ \\
0.96 & 12 & $0.883 \pm 0.060$ & $+0.0085$ \\
0.97 & 13 & $0.908 \pm 0.042$ & $+0.0246$ \\
0.98 & 16 & $0.912 \pm 0.039$ & $+0.0038$ \\
0.99 & 24 & $0.915 \pm 0.038$ & $+0.0029$ \\
\bottomrule
\end{tabular}
\end{table}

\begin{figure}[p]
    \centering
    \includegraphics[width=0.9\linewidth]
    {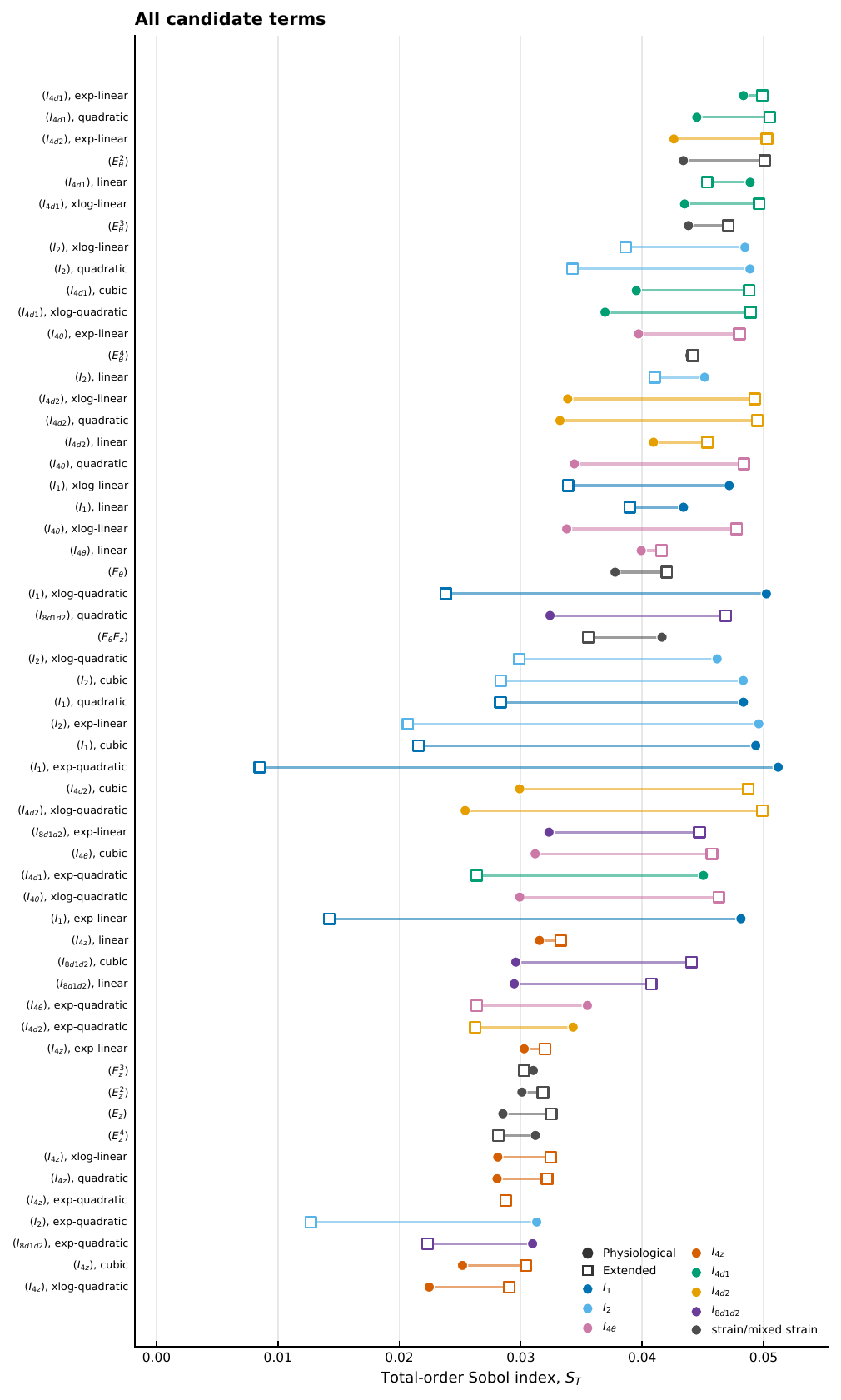}
    \caption{
    Total-order normalized Sobol indices, $S_T$, for all candidate strain-energy terms evaluated over the physiological and extended deformation grids. Filled circles and open squares denote the physiological and extended grids, respectively, and colors indicate the associated invariant or strain family.
    }
    \label{fig:supp_sobol_all_terms}
\end{figure}

\subsubsection{Near-Equivalent Constitutive Forms Across Individual Specimens}
\label{equivalent_const_forms}
To assess whether the final constitutive form was uniquely determined by the data, we evaluated all 320 compact candidate constitutive relations against the retained equation using an subject-specific near-equivalence criterion. A candidate was retained only if its combined RMSE was within \(5\%\) of the retained relation for every individual specimen. Under this stricter criterion, 16 equations were classified as near-equivalent across all nine subjects (Table \ref{tab:near-equivalent-equations-fungd}). These equations showed similar \(R^2\) and RMSE values despite differences in the nonlinear forms assigned to individual fiber-family terms, indicating that the shared constitutive architecture was robust, whereas the exact family-level basis functions were not uniquely identifiable.

\begin{landscape}
\begin{table}[t]
\centering
\caption{Candidate equations from the 320-candidate menu satisfying the subject-specific 5\% combined-RMSE criterion relative to the reduced model under the tied diagonal-family constraint ($d_1=d_2$; the $I_{4d1}$/$I_{4d2}$ responses share one coefficient). All equations include linear $I_1$, linear $I_2$, and the mixed strain term $E_\theta E_z$. Negative worst-case $\Delta$RMSE indicates improvement over the reference on every specimen.}
\label{tab:near-equivalent-equations-fungd}
\small
\begin{tabular}{r l r r r r r l l l l l}
\toprule
Rank & Eq. & \makecell{Mean\\RMSE} & \makecell{Median\\RMSE} & \makecell{Mean\\$R_w^2$} & \makecell{Median\\$R_w^2$} & \makecell{Worst\\$\Delta$RMSE (\%)} & $I_{4d1}$ & $I_{4d2}$ & $I_{4\mathrm{\theta}}$ & $I_{4\mathrm{z}}$ & $I_{8d1d2}$ \\
\midrule
1  & C292 & 4.062 & 3.861 & 0.9023 & 0.9073 & 0.18    & Exp Quad & Exp Quad & Cub      & Exp Lin  & Exp Lin \\
2  & C262 & 4.071 & 3.871 & 0.9019 & 0.9091 & 0.01    & Exp Quad & Exp Quad & xlog Lin & Exp Lin  & Exp Lin \\
3  & C272 & 4.081 & 3.864 & 0.9012 & 0.9091 & $-$0.16 & Exp Quad & Exp Quad & Quad     & Exp Lin  & Exp Lin \\
4  & C282 & 4.085 & 3.855 & 0.9012 & 0.9091 & $-$0.39 & Exp Quad & Exp Quad & Exp Quad & Exp Lin  & Exp Lin \\
5  & C302 & 4.087 & 3.874 & 0.9004 & 0.9072 & 0.22    & Exp Quad & Exp Quad & Exp Cub  & Exp Lin  & Exp Lin \\
6  & C281 & 4.092 & 3.879 & 0.9003 & 0.9093 & 0.22    & Exp Quad & Exp Quad & Exp Quad & Exp Lin  & Lin \\
7  & C271 & 4.099 & 3.856 & 0.9002 & 0.9093 & $-$0.36 & Exp Quad & Exp Quad & Quad     & Exp Lin  & Lin \\
8  & C261 & 4.101 & 3.868 & 0.9002 & 0.9093 & $-$0.05 & Exp Quad & Exp Quad & xlog Lin & Exp Lin  & Lin \\
9  & C280 & 4.101 & 3.935 & 0.9002 & 0.9079 & 4.76    & Exp Quad & Exp Quad & Quad     & Exp Quad & Exp Lin \\
10 & C301 & 4.103 & 3.862 & 0.8994 & 0.9075 & 0.08    & Exp Quad & Exp Quad & Exp Cub  & Exp Lin  & Lin \\
11 & C291 & 4.105 & 3.864 & 0.8994 & 0.9075 & 0.08    & Exp Quad & Exp Quad & Cub      & Exp Lin  & Lin \\
12 & C270 & 4.108 & 3.934 & 0.9001 & 0.9079 & 4.73    & Exp Quad & Exp Quad & xlog Lin & Exp Quad & Exp Lin \\
13 & C269 & 4.126 & 3.933 & 0.8990 & 0.9079 & 4.69    & Exp Quad & Exp Quad & xlog Lin & Exp Quad & Lin \\
14 & C279 & 4.131 & 3.933 & 0.8990 & 0.9079 & 4.71    & Exp Quad & Exp Quad & Quad     & Exp Quad & Lin \\
15 & C290 & 4.133 & 3.934 & 0.8979 & 0.9079 & 4.87    & Exp Quad & Exp Quad & Exp Quad & Exp Quad & Exp Lin \\
16 & C289 & 4.133 & 3.932 & 0.8986 & 0.9079 & 4.69    & Exp Quad & Exp Quad & Exp Quad & Exp Quad & Lin \\
\bottomrule
\end{tabular}
\end{table}
\end{landscape}

\subsubsection{Reduced SEF fits on remaining vein samples and best fit parameters}
The experimental data from all samples was fitted with the reduced model, and three representative samples shown in section \ref{results_final_model}. The remaining samples are shown in figures \ref{fig:supp-fits-vein2} and \ref{fig:supp-fits-vein3}.

\begin{figure}[t]
\centering
\includegraphics[width=\textwidth]{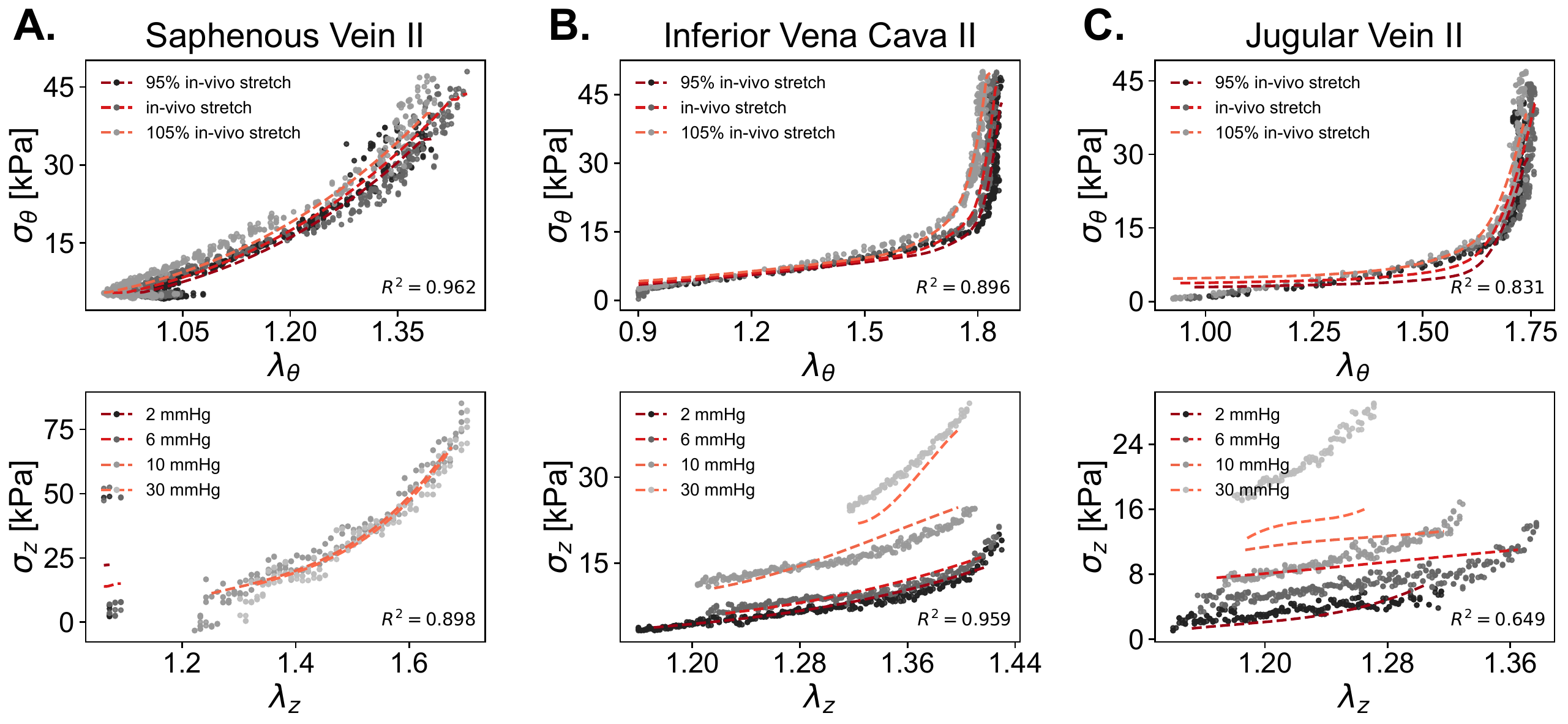}
\caption{Stress--stretch response and model fits for the second specimen of
each vein type (Saphenous Vein~II, Inferior Vena Cava~II, and Jugular
Vein~II), complementing the representative specimens shown in the main text.
Top row: circumferential Cauchy stress $\sigma_\theta$ versus circumferential
stretch $\lambda_\theta$ during pressure--diameter testing at 95\%, 100\%, and
105\% of the in-vivo axial stretch. Bottom row: axial Cauchy stress $\sigma_z$
versus axial stretch $\lambda_z$ during force--length testing at fixed luminal
pressures. Gray markers denote experimental data; dashed lines denote the
fitted reduced constitutive model with tied diagonal fiber families
($d_1=d_2$). Panel $R^2$ values are direction-specific coefficients of
determination.}
\label{fig:supp-fits-vein2}
\end{figure}

\begin{figure}[t]
\centering
\includegraphics[width=\textwidth]{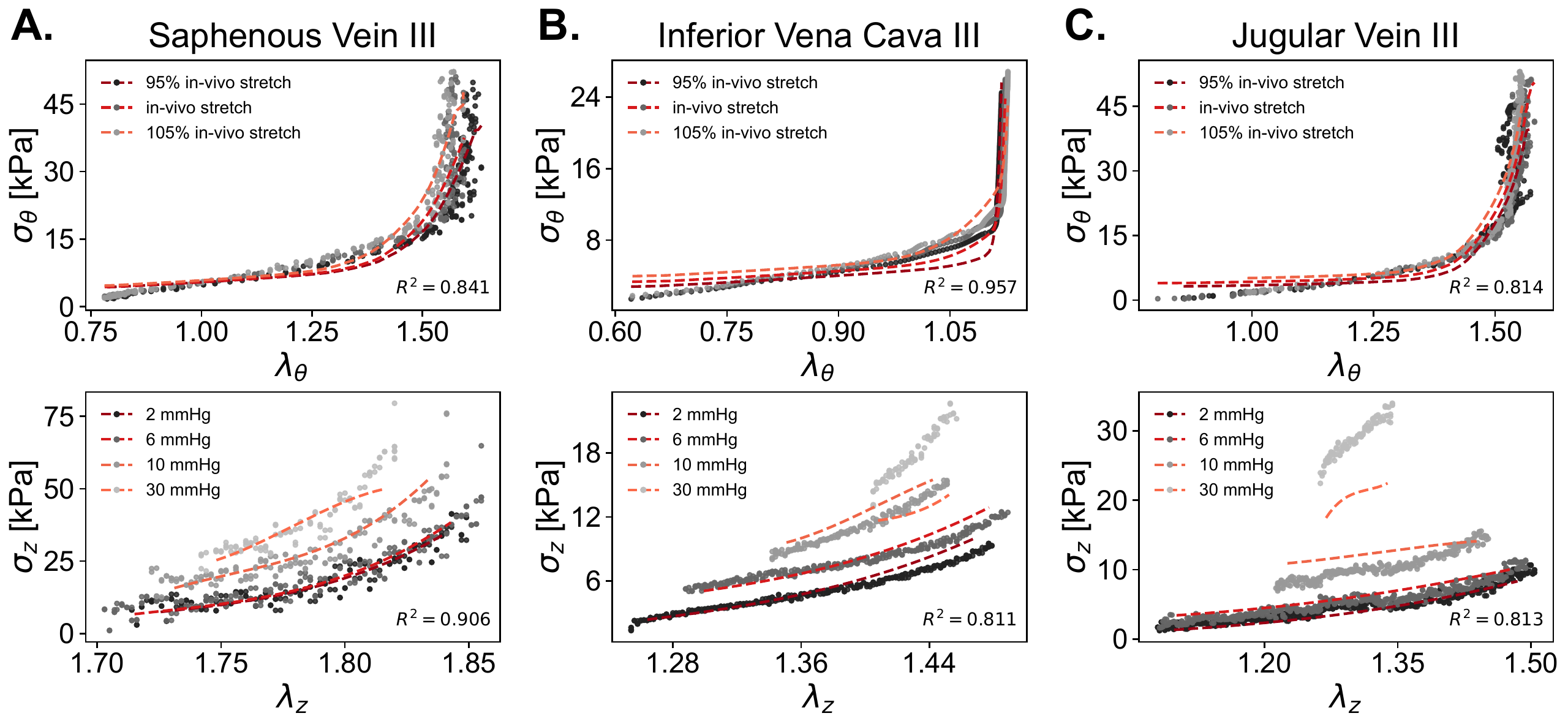}
\caption{Stress--stretch response and model fits for the third specimen of
each vein type (Saphenous Vein~III, Inferior Vena Cava~III, and Jugular
Vein~III), presented as in Fig.~\ref{fig:supp-fits-vein2}. Together with the
main-text figure, these panels cover all nine specimens used in the
constitutive model selection.}
\label{fig:supp-fits-vein3}
\end{figure}

\begin{table}[htbp]
\centering
\caption{Final best-fit material parameters of the final model.}
\label{tab:plaincubic_tied_d1d2_parameters}
\begin{tabular}{lrrrrrrr}
\hline
Sample & $c_1$ [kPa] & $c_2$ [kPa] & $c_{\theta}$ [kPa] & $c_d$ [kPa]& $c_z$ [kPa] & $c_{d1d2}$ [kPa]& $c_{\theta z}$ [kPa] \\
\hline
IVC1   & 0 & 0 & 58633.5 & 17.75 & 0.18 & 0.15 & 0.099 \\
IVC2   & 0 & 0 & 15.04 & 4812.05 & 0.060 & 0.091 & 0.047 \\
IVC3     & 0 & 0 & 0.71 & 320.25 & 0.011 & 0.058 & 0 \\
jugular1 & 0 & 0 & 35 & 0.024 & 0.0046 & 0.042 & 0 \\
jugular2 & 0 & 0.0052 & 5.34 & 45.89 & 0 & 0 & 0 \\
jugular3 & 0 & 0 & 18.10 & 50.52 & 0.044 & 0.012 & 0.14 \\
saphaneous1  & 0 & 0 & 4732.73 & 3.080 & 0.055 & 0 & 0.17 \\
saphaneous2  & 0 & 0 & 0.036 & 0.086 & 0.013 & 0 & 0 \\
saphaneous3  & 0.011 & 0 & 4.14 & $1.04 \times 10^{7}$ & $2.03 \times 10^{-6}$ & 0.038 & 0 \\
\hline
\end{tabular}
\end{table}

\end{appendices}

\end{document}